\documentclass[aip,jcp,amsmath,amssymb,reprint]{revtex4-1}

\usepackage{graphicx}
\usepackage{dcolumn}
\usepackage{bm}
\usepackage[utf8]{inputenc}
\usepackage[T1]{fontenc}
\usepackage{mathptmx}
\usepackage{etoolbox}
\usepackage{multirow}
\usepackage{longtable}
\usepackage[table]{xcolor}
\usepackage{colortbl}
\usepackage{makecell}
\usepackage{mathtools}
\usepackage{booktabs}
\usepackage{xcolor}
\usepackage{array}
\usepackage{placeins}
\usepackage{float}

\makeatletter
\def\@email#1#2{%
 \endgroup
 \patchcmd{\titleblock@produce}
  {\frontmatter@RRAPformat}
  {\frontmatter@RRAPformat{\produce@RRAP{*#1\href{mailto:#2}{#2}}}\frontmatter@RRAPformat}
  {}{}
}%
\makeatother

\begin{document}

\preprint{AIP/123-QED}

\title[Symmetry and Redundancy in Cremer--Pople Phases]{Symmetry and Redundancy in Cremer--Pople Phases for Ring Puckering}

\author{Anxo Lema-Saavedra}
\affiliation{
Centro Singular de Investigaci\'{o}n en Qu\'{i}mica Biol\'{o}xica e
Materiais Moleculares (CIQUS), C/ Jenaro de la Fuente s/n,\\
Universidade de Santiago de Compostela, 15782 Santiago de Compostela,
Spain}

\author{Antonio Fern\'{a}ndez-Ramos}
\email{qf.ramos@usc.es}
\affiliation{
Centro Singular de Investigaci\'{o}n en Qu\'{i}mica Biol\'{o}xica e
Materiais Moleculares (CIQUS), C/ Jenaro de la Fuente s/n,\\
Universidade de Santiago de Compostela, 15782 Santiago de Compostela,
Spain}
\affiliation{
Departamento de Qu\'{i}mica F\'{i}sica, Avda. das Ciencias s/n,\\
Universidade de Santiago de Compostela, 15782 Santiago de Compostela,
Spain}

\date{\today}

\begin{abstract}
Cremer--Pople (CP) coordinates provide an exact description of ring
puckering, yet direct sampling of their phases generates many
symmetry-equivalent representations of the same puckering pattern.
We develop a general analytical framework that identifies these
equivalences before Cartesian reconstruction and geometry optimization.
For homogeneous rings of arbitrary size, explicit transformations of
the CP phases and the signed special coordinate of even-membered rings
generate the complete symmetry orbit of the ordered elevation pattern.
We classify the redundancies into four types and derive a factorized
orbit-size formula linking CP multiplicity to puckering symmetry.
These results yield exact single-phase fundamental intervals and
sequentially reduced multiphase covering domains, including the
restriction to a positive special coordinate. The framework is extended
to rings containing one heteroatom or two identical heteroatoms, with
reductions determined by the operations that preserve their placement.
We also introduce a semidiscrete conformational nomenclature that
combines symbolic amplitude distributions with explicit active phases.
In an application to cyclohexadecane, symmetry reduction
alone decreases the number of sampled CP points by
approximately a factor of thirty. Combined with selective
phase-grid coarsening for three ordinary modes, the
reduction approaches two orders of magnitude.
These developments provide a systematic basis for generating,
classifying, and efficiently sampling ring puckering patterns in CP
space.
\end{abstract}
\maketitle

\section{Introduction}\label{sec:intro}

Flexible rings undergo large-amplitude puckering motions that cannot be
described by a single dihedral angle. Their conformation is instead a
collective property of all ring atoms, and its proper description has
been studied for more than half a century. A classic example is the
analysis of cyclopentane by Kilpatrick, Pitzer, and
Spitzer,~\cite{kilpatrick1947_cyclopentane_puckering} who showed that
the puckered forms of a five-membered ring can be described by just two
parameters, i.e., by an amplitude and by a phase angle, with interconversion taking
place through pseudorotation. This established the amplitude--phase
picture of ring puckering and strongly influenced later work on cyclic
conformational dynamics.

The description was later generalized to rings of arbitrary size
through the Cremer--Pople (CP)
coordinates.~\cite{cremer1975_cremer_pople_coordinates}
These coordinates define a unique mean plane and describe
the out-of-plane elevations \ensuremath{\{z_j\}} of the ring
atoms relative to this plane using \ensuremath{N-3}
independent puckering coordinates. These consist of an
amplitude and a phase for each ordinary mode and, for
even-membered rings, one additional signed coordinate
with no associated phase. For an $N$-membered ring, the CP amplitudes 
$q_m$ and phases $\phi_m$ are defined by~\cite{cremer1975_cremer_pople_coordinates}
\begin{equation}\label{eq:qm1}
q_m \cos \phi_m =
\sqrt{\frac{2}{N}} \sum_{j=1}^N z_j
\cos\left(\frac{2 \pi m (j - 1)}{N}\right),
\end{equation}
\begin{equation}\label{eq:qm2}
q_m \sin \phi_m =
-\sqrt{\frac{2}{N}} \sum_{j=1}^N z_j
\sin\left(\frac{2 \pi m (j - 1)}{N}\right),
\end{equation}
with $m=2,3,\ldots,M$, where $M=(N-1)/2$
if $N$ is odd and $M=N/2-1$ if $N$ is even.

If $N$ is even, there is one additional coordinate at $m=N/2$,
\begin{equation}\label{eq:qn2}
q_{N/2} =
\sqrt{\frac{1}{N}} \sum_{j=1}^N (-1)^{j - 1} z_j,
\end{equation}
which has no phase and may take either sign.

The inverse relation gives the elevations as functions of the amplitudes
and phases. For odd $N$,
\begin{equation}\label{eq:zjodd}
z_j = \sqrt{\frac{2}{N}} \sum_{m=2}^{M} q_m
\cos\!\left(\phi_m + \frac{2 \pi m (j - 1)}{N}\right),
\end{equation}
and for even $N$,
\begin{align}\label{eq:zjeven}
z_j &= \sqrt{\frac{2}{N}} \sum_{m=2}^{M} q_m
\cos\!\left(\phi_m + \frac{2 \pi m (j - 1)}{N}\right) \notag \\
&\quad + \sqrt{\frac{1}{N}}\, q_{N/2}(-1)^{j-1}.
\end{align}

Because CP coordinates describe the out-of-plane elevations of the
ring atoms, we examine the symmetry of this elevation pattern on a
reference regular polygon. This symmetry need not coincide with the
point-group symmetry of the full three-dimensional molecular structure. 
This is the same level at which Pickett and
Strauss~\cite{pickett1971_symmetry_conformation_cycloalkanes} analyzed
ring puckering, treating cycloalkane conformations as symmetry-adapted
out-of-plane displacements of a planar regular $N$-gon of $D_{Nh}$
symmetry. Their analysis classified the possible puckering modes,
counted the symmetry-related forms, identified low-barrier
pseudorotational paths, and derived selection rules for the associated
conformational dynamics.

Boeyens and Evans~\cite{boeyens1989_ring_group_theory} later established
the one-to-one correspondence between the symmetry-adapted coordinates
of Pickett and Strauss and the Cremer--Pople puckering parameters,
showing that the two formalisms are mathematically equivalent.
They also emphasized that the symmetry of the displacement pattern
\ensuremath{\{z_j\}} need not coincide with that of the full
three-dimensional molecular structure. More recently, Barone has incorporated local curvilinear
ring-puckering components into a general construction of
symmetry-adapted, non-redundant internal
coordinates.~\cite{Barone2026SONIC}

Previous studies have exploited CP symmetry in conformational
classification and sampling. Allen \emph{et al.}~\cite{allen1993_cycloheptane,allen1994_heteroatom,allen1996_cyclooctane}
partitioned the CP conformational spaces of cycloheptane and
cyclooctane into symmetry-equivalent asymmetric units.
Symmetry-based reductions have also been used for pyranose sugars
energy landscapes,~\cite{mayes2014_sugars_pucker} seven-membered-ring
metadynamics,~\cite{rovira2025_cycloheptane} and potential-energy
scans of four- to six-membered rings.~\cite{charvati2023}

Other studies improved sampling by restricting the explored
region of CP space~\cite{biarnes2007_glucopyranose,ardevol2010_mannopyranose}
or by using compact CP collective variables.~\cite{sega2009_puckering_fes,sega2011_pickett_cp}
These applications motivate a unified formulation of the
symmetry action in CP space and its use in constructing
reduced search domains for arbitrary ring size.

Our previous work~\cite{AFR2026rings} introduced a ring
reconstruction algorithm (RRA) for generating chemically
reasonable geometries from CP coordinates and performing
systematic conformational searches. However, symmetry
multiplicities were determined only \emph{a posteriori}
by comparing optimized structures. Identifying equivalent
CP representations before reconstruction and optimization
can avoid redundant calculations while retaining at least
one representative of every symmetry-inequivalent
puckering pattern.

In this work, we develop an analytical framework for rings of
arbitrary size that connects CP redundancy with puckering symmetry
and provides symmetry-reduced domains for conformational searches. 
First, we determine which symmetry
operations produce identical CP coordinates, establishing the
multiplicity and number of distinct CP images before Cartesian
reconstruction and geometry optimization. Second, we derive
symmetry-reduced phase domains for conformational searches,
retaining at least one representative of every
symmetry-inequivalent puckering pattern.
The underlying transformations relate CP representations through
cyclic relabeling, index reflection, or inversion of the elevations
through the mean plane; each resulting representation is referred
to as a \emph{CP image}.

The reductions are developed for unsubstituted cycloalkanes
and rings containing one heteroatom or two identical
heteroatoms, retaining only index operations that preserve
the atom-type pattern. Their use for other systems requires
preservation of chemical connectivity and specified
stereochemistry. For energy searches in an achiral environment,
fully mirror-related molecular geometries are included in
the same equivalence class because they have identical energies.

Specifically, we derive (i) explicit transformations of the CP
phases and, for even-membered rings, the signed special
coordinate; (ii) a closed-form orbit-size formula relating CP
multiplicity to puckering symmetry; and (iii) exact single-phase
fundamental intervals and sequentially reduced multiphase
covering domains with residual boundary redundancy.
Their integration with the reconstruction and search strategy
of Ref.~\onlinecite{AFR2026rings} is illustrated by a systematic
search for low-energy conformers of cyclohexadecane.

\section{Symmetry transformations of Cremer--Pople coordinates}
\label{sec:symm_search}

When a conformational search is carried out directly in CP space,
different phase points may encode the same puckering conformation
because of the symmetry of the reference regular polygon. The relevant
operations act on the ordered index set $j=1,\ldots,N$ and, when
appropriate, on the global sign of the elevations, so that the
resulting transformed CP images belong to the same equivalence class.
The goal of this section is to derive the explicit phase
transformations that generate all such equivalent images.

All transformations are derived using the original Cremer--Pople
orientation convention, in which the positive CP normal is the one from
which the ring atoms are viewed in clockwise order. Thus, whenever an
index transformation reverses this orientation, the transformed CP image
is re-expressed using the original reference normal.

Throughout this derivation, the out-of-plane elevations introduced
previously are treated as an ordered vector
\ensuremath{\mathbf z=(z_1,\ldots,z_N)}, expressed in the common
CP-normal orientation. This vector, rather than the full Cartesian
coordinates of the molecule, is the primary object of the analysis.
Accordingly, the labels shown in the figures denote the ordered CP
elevation indices, i.e., the positions of the components of
\ensuremath{\mathbf z}, rather than fixed chemical atom labels. 

\subsection{Indistinguishable atoms}

Let \ensuremath{\mathcal M=\{m_1,\ldots,m_r\}} denote the set
of active phase-bearing modes, i.e., those indices
\ensuremath{m\in\{2,\ldots,M\}} for which
\ensuremath{q_m\ne0}. The indices are ordered as
\ensuremath{m_1<\cdots<m_r}, and
\ensuremath{r=|\mathcal M|} is the number of active
phase-bearing modes.

For even-membered rings with a nonzero special coordinate,
the complete active-mode set is
\ensuremath{\mathcal M_{\pm}=\mathcal M\cup\{N/2\}},
with \ensuremath{r_{\pm}=r+1}.
When sampling is restricted to \ensuremath{q_{N/2}>0},
we denote this set by \ensuremath{\mathcal M_+}
and its size by \ensuremath{r_+=r+1}.
If the special coordinate is absent or zero, the complete
active-mode set is \ensuremath{\mathcal M}.

All operations that involve ring-index changes are
understood modulo \ensuremath{N}, and all phase
transformations are taken modulo \ensuremath{2\pi}. 
For indistinguishable ring atoms, the relevant
operations are: (i) cyclic index rotation, (ii) index reflection with orientation reversal, and
(iii) global inversion of the elevations.

\begin{figure}[!htpb]
  \centering
  \includegraphics[width=0.50\textwidth]{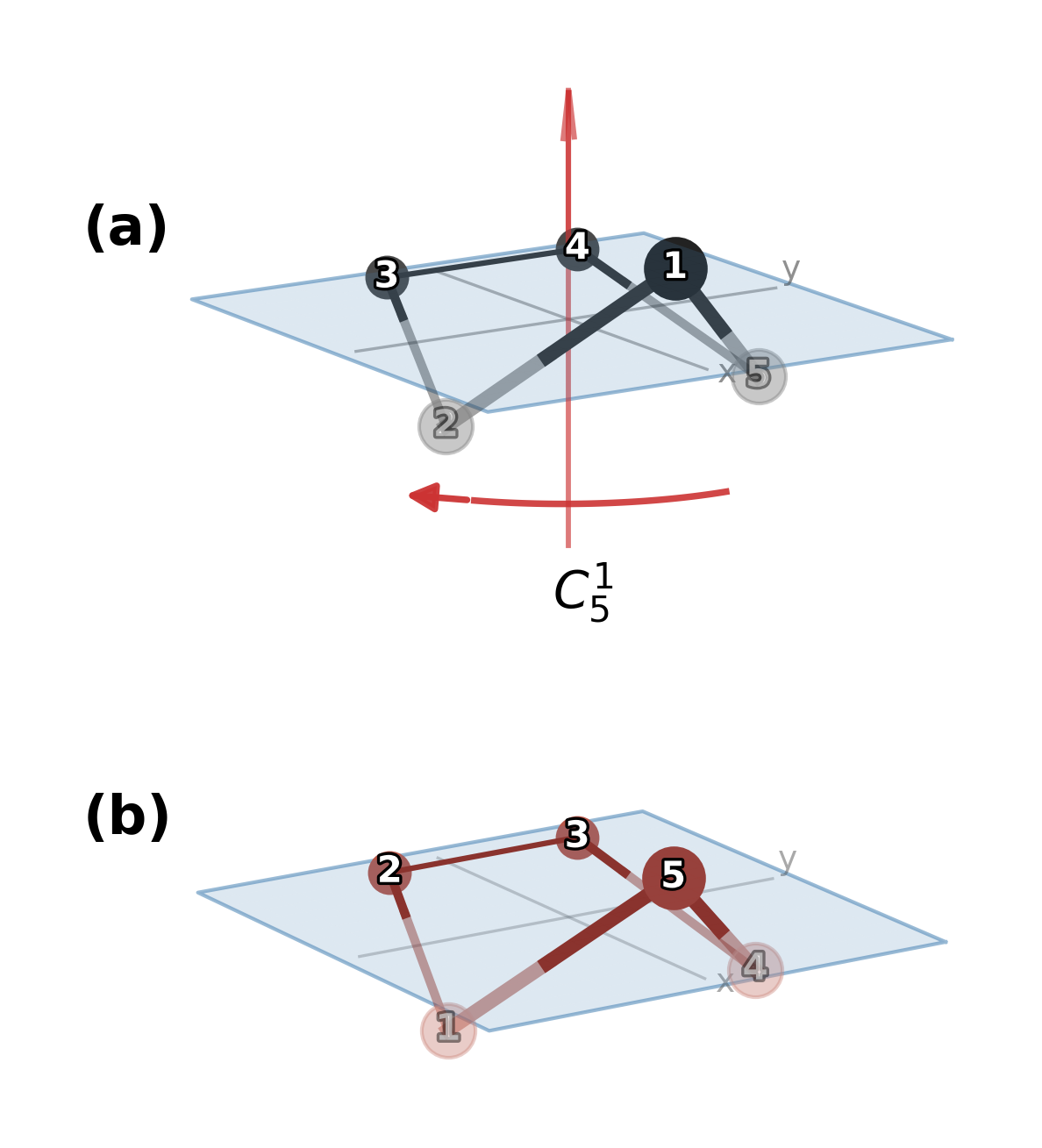}
\caption{Cyclic rotation of the ordered CP elevation vector for $N=5$:
(a) reference CP image, mean plane, and direction associated with the
$C_5^1$ transformation; (b) transformed CP image, in which the elevation
component assigned to position $j$ is taken from the original component
at position $j+1$.}
  \label{fig:c5_1}
\end{figure}

\paragraph*{(i) Cyclic rotation of indices.}
Let
\begin{equation}\label{eq:Rk_def}
\mathcal{R}_{k}: z_j \mapsto z'_j = z_{j+k},
\qquad k=0,1,\ldots,N-1,
\end{equation}
where the indices are taken modulo \(N\). In the present convention,
\(\mathcal{R}_k\) is a relabeling operation: the elevation assigned to
ring position \(j\) after the operation is the elevation that, before the
operation, belonged to position \(j+k\), as illustrated in
Fig.~\ref{fig:c5_1}.

Substituting Eq.~(\ref{eq:zjodd}) or
Eq.~(\ref{eq:zjeven}) into Eq.~(\ref{eq:Rk_def}),
the contribution from the phase-bearing modes becomes
\begin{equation}\label{eq:Rk_action}
z'_{j,\phi} = \sqrt{\frac{2}{N}} \sum_{m \in \mathcal{M}} q_m
\cos\!\left(\phi_m + \frac{2\pi m (j+k-1)}{N}\right),
\end{equation}
so that the transformed phase is
\begin{equation}\label{eq:Rk_phase}
\phi'_m = \phi_m + \frac{2\pi m k}{N}.
\end{equation}

Here $z'_{j,\phi}$ denotes the contribution from the phase-bearing
modes; for odd $N$, it is the complete transformed elevation.
If $N$ is even, the special coordinate transforms as
\begin{equation}\label{eq:Rk_qhalf}
q'_{N/2} = (-1)^k q_{N/2}.
\end{equation}
The factor $(-1)^k$ reflects the alternating character of the
$m=N/2$ mode; shifting the labeling by one ring position reverses its
sign.

\paragraph*{(ii) Index reflection with orientation reversal.}
Consider the index operation
\begin{equation}\label{eq:P_def}
\mathcal{C}_{\ell}: j \mapsto j' = N-j+\ell+2,
\qquad \ell = 0,1,\ldots,N-1,
\end{equation}
again modulo $N$. This operation reflects the ordered index set of the
reference polygon. The parameter $\ell$ specifies the position of the
reflection axis. For even $N$, even values of $\ell$ correspond to axes
through two opposite atoms, whereas odd values correspond to axes through
the midpoints of two opposite bonds. For odd $N$, each reflection axis
passes through one atom and the midpoint of the opposite bond; the fixed
atom satisfies $2j=\ell+2$ modulo $N$.

Because this reflection reverses the index orientation, the CP normal is
also reversed. Following the Cremer--Pople orientation convention stated
above, the transformed elevations are re-expressed with the reference CP
normal. Writing $j$ as the output index, the resulting action is
\begin{equation}\label{eq:P_z}
z'_j=-z_{N-j+\ell+2}.
\end{equation}
The minus sign in Eq.~(\ref{eq:P_z}) accounts for the reversal of the CP normal.
The transformed elevations are obtained by substituting the CP expansion into Equation~(\ref{eq:P_z}).
Absorbing the overall minus sign into the cosine using
$-\cos\theta=\cos(\theta+\pi)$ gives
\begin{equation}\label{eq:zjP_step3}
\begin{split}
z'_{j,\phi}
&=
\sqrt{\frac{2}{N}} \sum_{m \in \mathcal{M}} q_m
\cos\!\left[
-\frac{2\pi m(j-1)}{N}
+\frac{2\pi m\ell}{N}
+\phi_m
+\pi
\right].
\end{split}
\end{equation}

Using the evenness of the cosine, this expression becomes
\begin{equation}\label{eq:zjP}
\begin{split}
z'_{j,\phi}
&=
\sqrt{\frac{2}{N}} \sum_{m \in \mathcal{M}} q_m
\cos\!\left[
\frac{2\pi m(j-1)}{N}
-\frac{2\pi m\ell}{N}
-\phi_m
-\pi
\right].
\end{split}
\end{equation}
The comparison with the standard CP expansion gives
\begin{equation}\label{eq:P_phase}
\phi'_m =
-\phi_m-\frac{2\pi m\ell}{N}+\pi.
\end{equation}
Thus, reflected indexing reverses each phase and introduces an
axis-dependent shift, while the reversal of the CP normal contributes
the additional $\pi$ offset.

If $N$ is even, the special mode $m=N/2$ must be treated separately
because it has no associated phase variable. Since \ensuremath{N} is even, writing
\begin{equation}
N-j+\ell+1=(j-1)+(N-2j+\ell+2)
\end{equation}
gives
\begin{equation}
\begin{split}
(-1)^{N-j+\ell+1}
&= (-1)^{j-1}(-1)^{N-2j+\ell+2}\\
&= (-1)^{j-1}(-1)^\ell.
\end{split}
\end{equation}
Thus, including the minus sign induced by CP-normal reversal,
the transformed contribution becomes
\begin{equation}
z'_{j,N/2}
=
\frac{(-1)^{\ell+1}q_{N/2}}{\sqrt{N}}(-1)^{j-1},
\end{equation}
so that the special coordinate transforms as
\begin{equation}
q'_{N/2}=(-1)^{\ell+1}q_{N/2}.
\end{equation}
Therefore, for the phase-bearing modes, this normal
reversal is represented by a $\pi$ shift of the phases; for the
special mode, it is represented by a sign change of
the amplitude.

\paragraph*{(iii) Global inversion of the elevations.}
Global inversion reflects all atomic elevations through the mean plane.
This operation can be encoded by a sign parameter $s\in\{+1,-1\}$:
\begin{equation}\label{eq:Is_def}
\mathcal{I}_{s}: z'_j = s\,z_j,
\qquad s\in\{+1,-1\}.
\end{equation}
For $s=+1$ this is the identity, while for $s=-1$ it reverses the sign
of every elevation. Since $-\cos(\theta)=\cos(\theta+\pi)$, inversion
acts on the phase variables as
\begin{equation}\label{eq:Is_phase}
\phi'_m = \phi_m + \pi\,\frac{1-s}{2}.
\end{equation}

\begin{figure}[!htpb]
  \centering
  \includegraphics[width=0.50\textwidth]{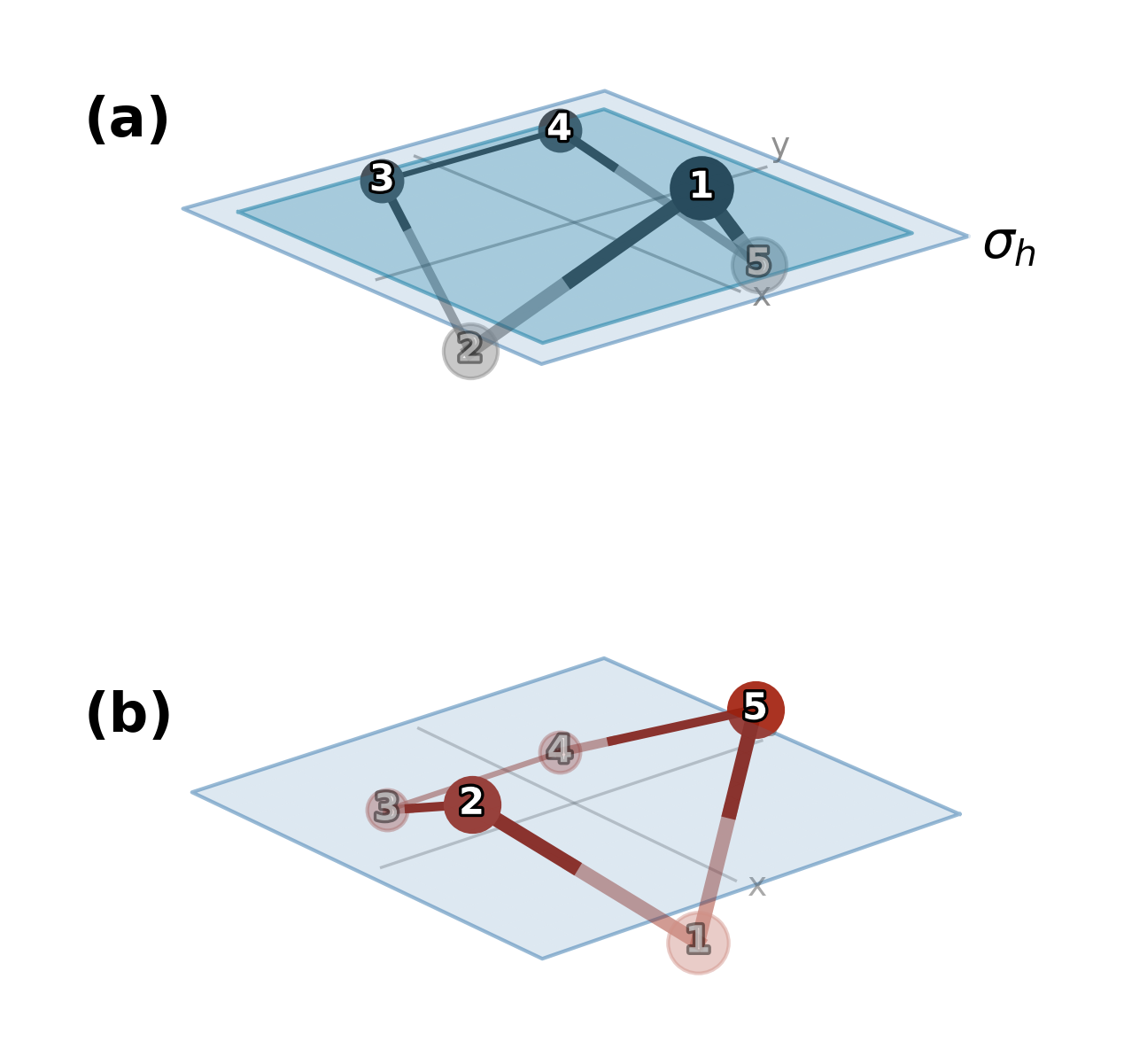}
\caption{Global inversion of the elevations for \(N=5\):
(a) shows the reference ring, mean plane, and the $\sigma_h$ plane which coincides with the mean plane;
(b) shows the ring after the inversion.}
  \label{fig:sigma_h}
\end{figure}

The effect of the global inversion is the same as applying a 
$\sigma_h$ operation over the geometry as shown in Fig.~\ref{fig:sigma_h}

For even $N$, the special coordinate transforms as
\begin{equation}\label{eq:Is_qn2}
q'_{N/2} = s\,q_{N/2}.
\end{equation}

Combining index reflection (when present), cyclic rotation, and
global inversion gives the general transformation law for the
phase variables:
\begin{equation}\label{eq:fullsymmetry_general}
\phi'_m = (-1)^p\,\phi_m
+ \frac{2\pi mk}{N}
- p\,\frac{2\pi m\ell}{N}
+ p\pi
+ \pi\,\frac{1-s}{2},
\end{equation}
where $p\in\{0,1\}$ distinguishes the rotation branch ($p=0$)
from the reflection branch ($p=1$), $k,\ell=0,\ldots,N-1$,
and $s\in\{\pm1\}$. The composition convention in
Eq.~(\ref{eq:fullsymmetry_general}) applies the index reflection
first, followed by the cyclic rotation and global inversion.
All phases are understood modulo $2\pi$.

In the reflection branch, varying \ensuremath{\ell}
at fixed \ensuremath{k=0}, or varying \ensuremath{k}
at fixed \ensuremath{\ell}, generates the same set of
index reflections. Using \ensuremath{N} rotations
and \ensuremath{N} reflections, each with two choices
of \ensuremath{s}, gives \ensuremath{4N} formal operations.

For even $N$, the special coordinate transforms as
\begin{equation}\label{eq:qn2_combined}
q'_{N/2}
=
s\,(-1)^{k+p(\ell+1)}\,q_{N/2}.
\end{equation}
The cyclic permutation contributes $(-1)^k$, while reflected
indexing contributes $(-1)^\ell$ when $p=1$. The additional
factor $(-1)^p$ accounts for the reversal of the CP normal,
and $s$ represents the independent global inversion.

These transformations can also be associated with the conventional
symmetry-operation labels of the regular reference polygon.
The correspondence between the parameters
\ensuremath{(k,p,\ell,s)} and the \ensuremath{D_{Nh}} labels is
listed in Tables~\ref{tbl:chartable_odd} and~\ref{tbl:chartable_even}.
In both tables, \ensuremath{\ell} is irrelevant in the rotation
branch, and \ensuremath{k=0} is used in the reflection branch
to avoid duplicate parametrizations.

\begin{table}[!htpb]
	\caption{Parameters \ensuremath{(k,p,\ell,s)}, induced phase mapping
		\ensuremath{\phi_m'}, and corresponding \ensuremath{D_{Nh}} labels for
		odd \ensuremath{N}, where \ensuremath{\kappa_m=2\pi m/N}.
		Reflection rows use \ensuremath{k=0}.
		All phase mappings are understood modulo \ensuremath{2\pi}.}
	\label{tbl:chartable_odd}
	\centering
	\renewcommand{\arraystretch}{1.6}
	\begin{tabular}{c @{\hspace{1.2em}} c @{\hspace{1.2em}} c l l}
		\hline\hline
		\ensuremath{p} & \ensuremath{s} & \ensuremath{k} or \ensuremath{\ell}
		& \ensuremath{\phi_m'} & \ensuremath{D_{Nh}}\\
		\hline
		\ensuremath{0} & \ensuremath{+1} & \ensuremath{0}
		& \ensuremath{\phi_m}
		& \ensuremath{E} \\[2pt]
		\ensuremath{0} & \ensuremath{+1} & \ensuremath{k=1,\ldots,N-1}
		& \ensuremath{\phi_m+\kappa_m k}
		& \ensuremath{C_N^k} \\[4pt]
		\ensuremath{0} & \ensuremath{-1} & \ensuremath{0}
		& \ensuremath{\phi_m+\pi}
		& \ensuremath{\sigma_h} \\[2pt]
		\ensuremath{0} & \ensuremath{-1} & \ensuremath{k=1,\ldots,N-1}
		& \ensuremath{\phi_m+\kappa_m k+\pi}
		& \ensuremath{S_N^k}\footnotemark[1] \\[4pt]
		\ensuremath{1} & \ensuremath{+1} & \ensuremath{\ell=0,\ldots,N-1}
		& \ensuremath{-\phi_m-\kappa_m\ell+\pi}
		& \ensuremath{C_2^{\prime,\ell}} \\[4pt]
		\ensuremath{1} & \ensuremath{-1} & \ensuremath{\ell=0,\ldots,N-1}
		& \ensuremath{-\phi_m-\kappa_m\ell}
		& \ensuremath{\sigma_v^{\ell}} \\[4pt]
		\hline\hline
	\end{tabular}
	\footnotetext[1]{Throughout this work, \ensuremath{S_N^k} denotes
		the composite operation \ensuremath{C_N^k\sigma_h}.
		The superscript specifies the cyclic rotation accompanying
		reflection through the reference mean plane.}
\end{table}

\begin{table}[!htpb]
\caption{Parameters $(k,p,\ell,s)$, induced phase mapping
$\phi_m'$, transformation of $q_{N/2}$, and 
$D_{Nh}$ label for even $N$, where $\kappa_m=2\pi m/N$. The notation \ensuremath{S_N^k}
follows the convention specified in Table~\ref{tbl:chartable_odd}.
Reflection rows use $k=0$. The special-coordinate mappings
include the effect of CP-normal reversal. All phase mappings
are understood modulo $2\pi$.}
\label{tbl:chartable_even}
\centering
\renewcommand{\arraystretch}{1.5}
\setlength{\tabcolsep}{4pt}
\small
\begin{tabular}{c c c l c l}
\hline\hline
$p$ & $s$ & $k$ or $\ell$ & $\phi_m'$
  & $q'_{N/2}$ & label \\
\hline
$0$ & $+1$ & $0$
  & $\phi_m$
  & $+q_{N/2}$ & $E$ \\[2pt]
$0$ & $+1$ & $k=1,\ldots,N-1$
  & $\phi_m+\kappa_m k$
  & $(-1)^k q_{N/2}$ & $C_N^k$ \\[2pt]
$1$ & $+1$ & $\ell=0,2,\ldots,N-2$
  & $-\phi_m-\kappa_m\ell+\pi$
  & $-q_{N/2}$ & $C_2^{\prime,\ell}$ \\[2pt]
$1$ & $+1$ & $\ell=1,3,\ldots,N-1$
  & $-\phi_m-\kappa_m\ell+\pi$
  & $+q_{N/2}$ & $C_2^{\prime\prime,\ell}$ \\[2pt]
$0$ & $-1$ & $0$
  & $\phi_m+\pi$
  & $-q_{N/2}$ & $\sigma_h$ \\[2pt]
$0$ & $-1$ & $k=1,\ldots,N-1$
  & $\phi_m+\kappa_m k+\pi$
  & $-(-1)^k q_{N/2}$ & $S_N^k$ \\[2pt]
$1$ & $-1$ & $\ell=0,2,\ldots,N-2$
  & $-\phi_m-\kappa_m\ell$
  & $+q_{N/2}$ & $\sigma_v^{\ell}$ \\[2pt]
$1$ & $-1$ & $\ell=1,3,\ldots,N-1$
  & $-\phi_m-\kappa_m\ell$
  & $-q_{N/2}$ & $\sigma_d^{\ell}$ \\[2pt]
\hline\hline
\end{tabular}
\end{table}

\subsection{Symmetry-transformed and Cartesian-derived CP coordinates}
\label{subsection:cp_normal}

The transformations derived above act directly on the CP representation,
that is, on the ordered elevation vector and on the corresponding CP
coordinates. In the convention used throughout this work, the ring labels
denote CP elevation indices, i.e. the positions of the components of
$\mathbf z=(z_1,\ldots,z_N)$, rather than material atom labels followed
through a Cartesian transformation.

\begin{figure}[!htpb]
  \centering
  \includegraphics[width=0.50\textwidth]{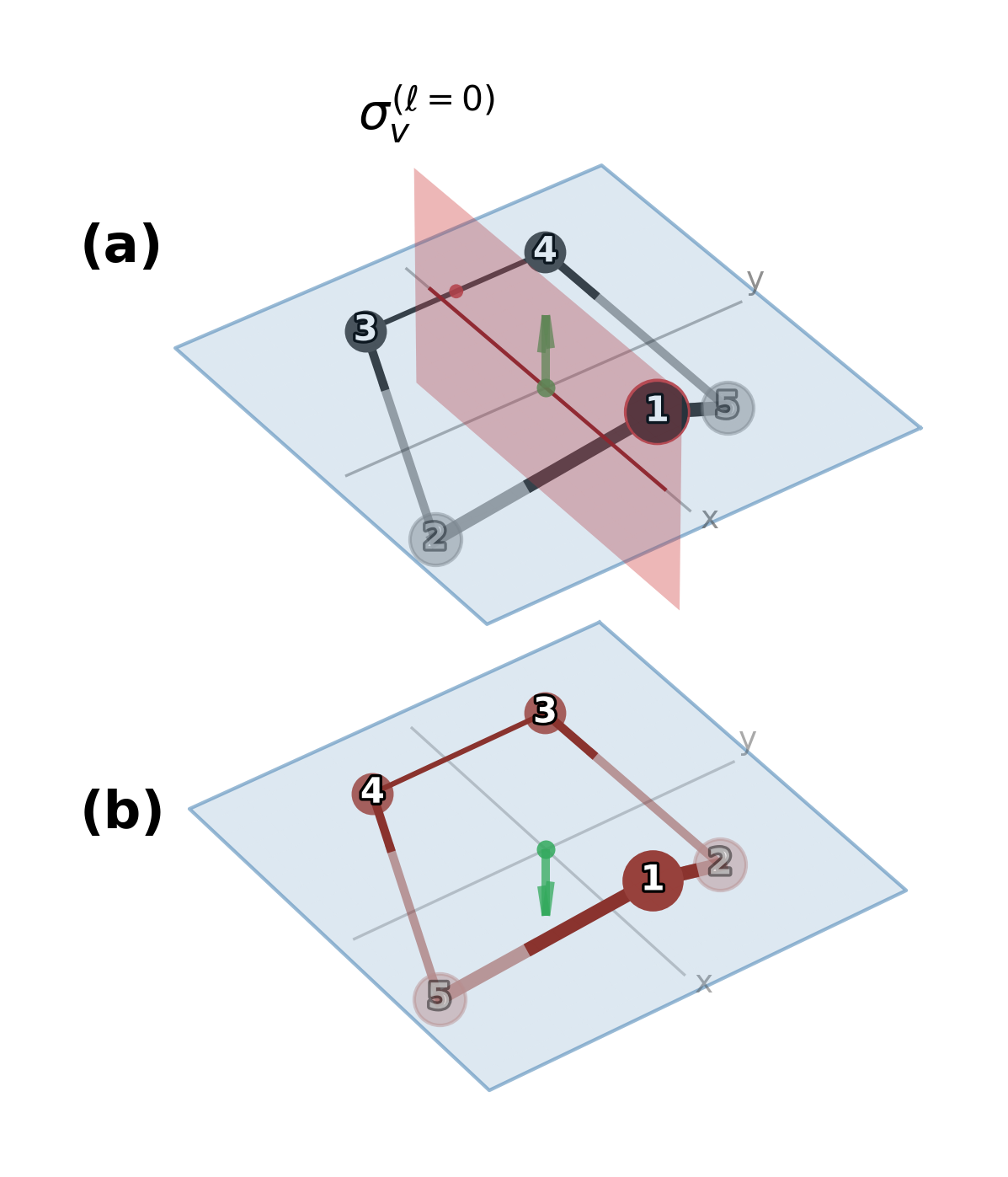}
  \caption{Reflected indexing with orientation reversal for an envelope
  conformation of cyclopentane: (a) reference CP image, mean plane,
  vertical reflection plane $\sigma_v^{\ell=0}$, and oriented CP
  normal in green; (b) reflected CP indexing, which reverses the spatial
  cyclic order associated with the Cartesian representation.}
  \label{fig:sigma_v}
\end{figure}

This distinction becomes important when CP coordinates are recomputed from
a transformed Cartesian geometry. This situation is illustrated in Fig.~\ref{fig:sigma_v} for the vertical
reflection $\sigma_v^{\ell=0}$ of an envelope conformation of
cyclopentane. In the reference CP image [Fig.~\ref{fig:sigma_v}(a)], the
ordered CP indices define the positive CP normal. After the reflected-index
operation [Fig.~\ref{fig:sigma_v}(b)], two equivalent readings are possible,
depending on what the labels are taken to represent. If the labels denote
CP elevation indices, i.e. fixed positions in the ordered vector
$\mathbf z$, the reflected image is projected using the fixed CP-index
convention and the recomputed CP coordinates coincide with the analytical
transformation. In this convention, the original ordered vector of
elevations $(z_1,z_2,z_3,z_4,z_5)$ is transformed by
$\sigma_v^{\ell=0}$ into $(z_1,z_5,z_4,z_3,z_2)$. 

By contrast, if the
labels are followed as material atom labels in the Cartesian ring file,
the same reflected geometry is read with the opposite cyclic orientation.
This reindexing reverses the CP normal defined by the ordered atoms.
Consequently, an additional $180^\circ$ phase shift for the phase-bearing
modes must be included in order to express the Cartesian-derived CP
coordinates in the same convention as the analytical expression.
For even-membered rings, the same change of normal
also reverses the sign of the special coordinate.

This shift is purely conventional because once the
reflected structure is expressed again in the fixed CP-index convention,
or with the same positive CP normal as the reference structure, the
Cartesian-derived and analytically transformed CP coordinates are
consistent.

\subsection{Rings with heteroatoms}\label{sec:heteroatom}

When distinguishable atoms (nodes in the ring) are present, the allowed symmetry
transformations are restricted to those that preserve the heteroatom
pattern, either by fixing each heteroatom or by exchanging identical
ones. As in the homogeneous case, the resulting equivalence is defined
at the level of the CP image, not necessarily at the level of the full
three-dimensional molecular geometry. Global inversion $\mathcal{I}_s$
remains unrestricted but cyclic rotations are allowed only if they preserve the heteroatom pattern.

\paragraph*{One heteroatom.}
If one node $j_a$ differs from the rest, cyclic rotations
$\mathcal{R}_k$ are restricted to the identity $k=0$. 
A reflection $\mathcal{C}_{\tilde{\ell}}$ is allowed only
if it fixes the special node, which requires
$\tilde{\ell}\equiv2(j_a-1)\pmod N$.
The corresponding axis passes through the heteroatom.
For even $N$, its axis index is necessarily even.
The allowed operations therefore form a
four-element group of $C_{2v}$ type in puckering space, generated by
$\{E, \sigma_v, \sigma_h, C_2\}$. All nontrivial cyclic rotations are
forbidden because they move the heteroatom to a different ring
position, producing a distinct labeled puckering conformation. Only the four
operations that preserve the heteroatom site remain redundant. 
The allowed phase
mappings are listed in Table~\ref{tbl:chartable_hetero_c2v}.

\paragraph*{Two identical heteroatoms.}
If two nodes $j_a<j_b$ are identical to each other but distinct from
the rest, the allowed symmetry operations depend on their relative
placement.

Two types of reflection $\mathcal{C}_{\tilde{\ell}}$ are possible:
\begin{itemize}
  \item \emph{Swapping axis}
    ($\mathcal{C}_{\tilde{\ell}}(j_a)=j_b$,
     $\mathcal{C}_{\tilde{\ell}}(j_b)=j_a$):
    always allowed, with
    \begin{equation}\label{eq:ell_swap}
      \tilde{\ell} = j_a + j_b - 2.
    \end{equation}
For even $N$, an even value of $\tilde{\ell}$ corresponds
to an axis through two opposite atoms, whereas an odd value
corresponds to an axis through the midpoints of two opposite
bonds. For odd $N$, every reflection axis passes through
one atom and the midpoint of the opposite bond.
  \item \emph{Fixing axis}
    ($\mathcal{C}_{\tilde{\ell}}(j_a)=j_a$,
     $\mathcal{C}_{\tilde{\ell}}(j_b)=j_b$):
    requires $j_b = j_a + N/2$ (antipodal placement, even $N$ only),
    with $\tilde{\ell}_1 = 2(j_a - 1)$.
\end{itemize}

When the two heteroatoms are \emph{not} antipodal
($j_b - j_a \neq N/2$), only the swapping axis is available and
rotations are restricted to $k=0$. The residual action on the CP image
again has $C_{2v}$ type, with $\tilde{\ell} = j_a + j_b - 2$; the
phase mappings take the same form as in the one-heteroatom case and are
given by Table~\ref{tbl:chartable_hetero_c2v}, with $\tilde{\ell}$ now
defined by Equation~(\ref{eq:ell_swap}).

\begin{table}[!htpb]
\caption{Generator parameters $(p,s)$, induced phase mapping $\phi_m'$,
and corresponding $C_{2v}$ operation for heteroatom-containing rings,
with fixed reflection-axis index $\tilde{\ell}$ and $k=0$.}
\label{tbl:chartable_hetero_c2v}
\centering
\renewcommand{\arraystretch}{1.6}
\begin{tabular}{c @{\hspace{1.2em}} c @{\hspace{1.2em}} l l}
\hline\hline
$p$ & $s$ &  $\phi_m'$ & $C_{2v}$\\
\hline
$0$ & $+1$
  & $\phi_m$
  & $E$ \\[2pt]
$1$ & $+1$
  & $-\phi_m - \kappa_m\tilde{\ell} + \pi$
  & $C_2$ \\[4pt]
$0$ & $-1$
  & $\phi_m + \pi$
  & $\sigma_h$ \\[2pt]
$1$ & $-1$ 
  & $-\phi_m - \kappa_m\tilde{\ell}$
  & $\sigma_v$ \\[4pt]
\hline\hline
\end{tabular}
\end{table}

\begin{table}[!htpb]
\caption{Generator parameters, phase mapping
\ensuremath{\phi_m'}, transformation of the special coordinate
\ensuremath{q_{N/2}}, and reference-subgroup labels for a ring
with two identical heteroatoms at antipodal positions.
Here \ensuremath{\kappa_m=2\pi m/N},
\ensuremath{\tilde{\ell}_1=2(j_a-1)}, and
\ensuremath{\tilde{\ell}_2=\tilde{\ell}_1+N/2}, with axis indices
understood modulo \ensuremath{N}. All phase mappings are
understood modulo \ensuremath{2\pi}.}
\label{tbl:chartable_hetero2_antipodal}
\centering
\renewcommand{\arraystretch}{1.5}
\setlength{\tabcolsep}{4pt}
\small
\begin{tabular}{c c c l c l}
\hline\hline
\ensuremath{p} & \ensuremath{s}
& \ensuremath{k,\tilde{\ell}}
& \ensuremath{\phi_m'}
& \ensuremath{q_{N/2}'}
& Label \\
\hline
\ensuremath{0} & \ensuremath{+1} & \ensuremath{0,--}
  & \ensuremath{\phi_m}
  & \ensuremath{+q_{N/2}}
  & \ensuremath{E} \\[2pt]

\ensuremath{0} & \ensuremath{+1} & \ensuremath{N/2,--}
  & \ensuremath{\phi_m+\pi m}
  & \ensuremath{(-1)^{N/2}q_{N/2}}
  & \ensuremath{C_2} \\[2pt]

\ensuremath{1} & \ensuremath{+1}
  & \ensuremath{0,\tilde{\ell}_1}
  & \ensuremath{-\phi_m-\kappa_m\tilde{\ell}_1+\pi}
  & \ensuremath{-q_{N/2}}
  & \ensuremath{C_2'} \\[2pt]

\ensuremath{1} & \ensuremath{+1}
  & \ensuremath{0,\tilde{\ell}_2}
  & \ensuremath{-\phi_m-\kappa_m\tilde{\ell}_2+\pi}
  & \ensuremath{(-1)^{\tilde{\ell}_2+1}q_{N/2}}
  & \ensuremath{C_2''} \\[2pt]

\ensuremath{0} & \ensuremath{-1} & \ensuremath{0,--}
  & \ensuremath{\phi_m+\pi}
  & \ensuremath{-q_{N/2}}
  & \ensuremath{\sigma_h} \\[2pt]

\ensuremath{0} & \ensuremath{-1} & \ensuremath{N/2,--}
  & \ensuremath{\phi_m+\pi m+\pi}
  & \ensuremath{(-1)^{N/2+1}q_{N/2}}
  & \ensuremath{i} \\[2pt]

\ensuremath{1} & \ensuremath{-1}
  & \ensuremath{0,\tilde{\ell}_1}
  & \ensuremath{-\phi_m-\kappa_m\tilde{\ell}_1}
  & \ensuremath{+q_{N/2}}
  & \ensuremath{\sigma_v} \\[2pt]

\ensuremath{1} & \ensuremath{-1}
  & \ensuremath{0,\tilde{\ell}_2}
  & \ensuremath{-\phi_m-\kappa_m\tilde{\ell}_2}
  & \ensuremath{(-1)^{\tilde{\ell}_2}q_{N/2}}
  & \ensuremath{\sigma_v'} \\[2pt]
\hline\hline
\end{tabular}
\end{table}

When the two heteroatoms are \emph{antipodal}
($j_b = j_a + N/2$, even $N$ only), both reflection axes are available
and the rotation $k=N/2$, which exchanges
$j_a \leftrightarrow j_b$, is also allowed. The residual action on the
CP image therefore extends to an eight-element group of
$D_{2h}$ type. Here the rotation $k=N/2$ is allowed because it
exchanges the two identical heteroatoms. The
full set of phase mappings is given in
Table~\ref{tbl:chartable_hetero2_antipodal}. 
Compared to the $C_{2v}$ case, the additional operations are the
antipodal rotation $C_2$, the swapping reflection $\sigma_v'$,
inversion $i$, and the second in-plane twofold operation $C_2''$.

The labels \ensuremath{C_2'} and \ensuremath{C_2''} distinguish
the two mutually perpendicular in-plane axes of this
\ensuremath{D_{2h}}-type subgroup: the first passes through the
heteroatoms, whereas the second exchanges them. Likewise,
\ensuremath{\sigma_v} and \ensuremath{\sigma_v'} denote the
vertical planes containing these axes. These are local subgroup
labels and need not coincide with the prime and double-prime
conventions of the parent \ensuremath{D_{Nh}} group.

\section{Redundancy, orbit reduction, and puckering symmetry}
\label{sec:orbit_size}

The transformations derived in Sec.~\ref{sec:symm_search} generate
\ensuremath{4N} formal CP images of a homogeneous ring. Some of these
images may coincide because different operations produce the same
ordered elevation pattern. The collection of distinct images is
called the CP orbit.

The orbit is determined by counting the operations
that reproduce the reference elevation pattern,
\ensuremath{\mathbf z=(z_1,\ldots,z_N)}. Such operations leave all
active CP coordinates unchanged, including the signed coordinate
\ensuremath{q_{N/2}} for even-membered rings. Phases associated with
zero amplitudes are disregarded because they do not affect the
elevation pattern.

We denote the number of operations that reproduce the reference
pattern by \ensuremath{m_{\rm CP}}. This number is also the
multiplicity of each distinct CP image. Therefore, for a homogeneous ring the CP orbit is,
\begin{equation}
|\mathcal O(\{\phi_m\})|
=
\frac{4N}{m_{\rm CP}}.
\label{eq:orbit_stabilizer_simple}
\end{equation}
If only the identity reproduces the reference pattern,
\ensuremath{m_{\rm CP}=1} and all \ensuremath{4N} images are
distinct. Each additional operation that leaves the pattern
unchanged is considered redundant and reduces the number of
distinct images.

\subsection{Redundancy types and orbit size}

The redundancies can be classified into four types:

\begin{itemize}
  \item Type-R: two cyclic shifts produce the same CP image.
  These redundancies correspond to rotations about the axis
  normal to the reference plane.

  \item Type-C: reflected indexing with normal reversal produces
  the same CP image as a cyclic shift. These redundancies
  correspond to twofold rotations  about axes lying in the
  reference plane ($C_{2,\perp}$).

  \item Type-I: global inversion of the elevations produces a
  CP image already obtained by a cyclic shift. These redundancies
  correspond to cyclic shifts combined with elevation inversion.

  \item Type-P: reflected indexing combined with global inversion
  produces the same CP image as a cyclic shift. These redundancies
  correspond to mirror-plane operations.
\end{itemize}

The conditions for each type are derived in
Appendix~\ref{app:cp_redundancy}. Type-R and Type-I depend exclusively on
the active modes and are phase independent, whereas Type-P and Type-C also depend on the
phase values.

We denote the rotational reduction factor by \ensuremath{f_{\rm R}}.
The other three factors are denoted as \ensuremath{f_{\rm I}}, \ensuremath{f_{\rm P}}, and
\ensuremath{f_{\rm C}}. They are two when the corresponding redundancy condition
is satisfied and one otherwise. These contributions are not
independent and combining operations from any two of the types
P, I, and C produces an operation of the third type which is dependent on the other two.
Therefore, when all three types are present, multiplying their factors
counts one reduction twice. To correct this repeated counting, we introduce the overlap factor
\ensuremath{g}, defined as
\begin{equation}
g=
\begin{cases}
2, & f_{\rm P}=f_{\rm I}=f_{\rm C}=2,\\
1, & \text{otherwise}.
\end{cases}
\label{eq:degeneracy_g}
\end{equation}
The CP multiplicity is then
\begin{equation}
m_{\rm CP}
=
\frac{f_{\rm R}f_{\rm P}f_{\rm I}f_{\rm C}}{g},
\label{eq:mcp_factorized}
\end{equation}
and the CP orbit becomes
\begin{equation}
|\mathcal O(\{\phi_m\})|
=
g\frac{4N}{f_{\rm R}f_{\rm P}f_{\rm I}f_{\rm C}}.
\label{eq:orbit_size_revised}
\end{equation}

Equation~(\ref{eq:orbit_size_revised}) applies to the full set of
formal operations of a homogeneous ring. For rings containing
heteroatoms, only operations that preserve the heteroatom pattern
are allowed. The number of distinct CP images is then obtained
by dividing the number of allowed operations by the number of
those operations that reproduce the reference CP pattern.

\subsection{CP multiplicities}

There is a connection  between CP
redundancy and puckering symmetry. More generally, the operations
that leave an elevation pattern unchanged determine its symmetry
group, while their number gives the CP multiplicity and hence the
orbit size. Table~\ref{tab:ring_cp_examples} summarizes this
correspondence for representative homogeneous ring patterns.
The point-group labels describe the elevation pattern on the
regular reference polygon; they also apply to an ideal reconstructed
ring when its in-plane geometry is compatible with these operations,
but need not describe an arbitrary Cartesian molecular geometry.

\begin{table}[!htbp]
\centering
\caption{Representative homogeneous ring patterns, their symmetry
groups, CP reduction factors, multiplicities, and orbit sizes.
The symmetry labels refer to the elevation patterns on the regular
reference polygon.}
\label{tab:ring_cp_examples}
\renewcommand{\arraystretch}{1.25}
\setlength{\tabcolsep}{3pt}
\small
\begin{tabular}{llccccccc}
\hline\hline
Pattern & Group
& \ensuremath{f_{\rm R}}
& \ensuremath{f_{\rm C}}
& \ensuremath{f_{\rm I}}
& \ensuremath{f_{\rm P}}
& \ensuremath{g}
& \ensuremath{m_{\rm CP}}
& \ensuremath{|\mathcal O|} \\
\hline
Five-membered, generic
& \ensuremath{C_1}
& 1 & 1 & 1 & 1 & 1 & 1 & 20 \\
Five-membered, envelope
& \ensuremath{C_s}
& 1 & 1 & 1 & 2 & 1 & 2 & 10 \\
Five-membered, twist
& \ensuremath{C_2}
& 1 & 2 & 1 & 1 & 1 & 2 & 10 \\
Six-membered, chair
& \ensuremath{D_{3d}}
& 3 & 2 & 2 & 2 & 2 & 12 & 2 \\
Six-membered, boat
& \ensuremath{C_{2v}}
& 2 & 1 & 1 & 2 & 1 & 4 & 6 \\
Six-membered, skew
& \ensuremath{D_2}
& 2 & 2 & 1 & 1 & 1 & 4 & 6 \\
Six-membered, half-chair
& \ensuremath{C_2}
& 1 & 2 & 1 & 1 & 1 & 2 & 12 \\
Six-membered, half-boat
& \ensuremath{C_s}
& 1 & 1 & 1 & 2 & 1 & 2 & 12 \\
\ensuremath{C_{\rm N}} puckered, \ensuremath{S_{2n}}
& \ensuremath{S_{2n}}
& \ensuremath{n} & 1 & 2 & 1 & 1
& \ensuremath{2n}
& \ensuremath{2N/n} \\
\ensuremath{C_{\rm N}} puckered, \ensuremath{D_n}
& \ensuremath{D_n}
& \ensuremath{n} & 2 & 1 & 1 & 1 & 2n
& \ensuremath{2N/n} \\
\ensuremath{C_{\rm N}} puckered, \ensuremath{D_{nd}}
& \ensuremath{D_{nd}}
& \ensuremath{n} & 2 & 2 & 2 & 2 & 4n
& \ensuremath{N/n} \\
\ensuremath{N}-membered, planar
& \ensuremath{D_{Nh}}
& \ensuremath{N} & 2 & 2 & 2 & 2
& \ensuremath{4N} & 1 \\
\hline\hline
\end{tabular}
\end{table}

For example, cyclopentane has a single phase-bearing mode,
\ensuremath{m=2}, with neither Type-R nor Type-I redundancy.
A generic phase therefore produces twenty distinct CP images.
At an envelope phase, such as \ensuremath{\phi_2=0^\circ},
Type-P reduces the orbit through a mirror-type operation
(\ensuremath{C_s} symmetry). At a twist phase, such as
\ensuremath{\phi_2=18^\circ}, Type-C gives the same reduction
through an in-plane twofold rotation
(\ensuremath{C_2} symmetry). In both cases,
\ensuremath{g=1}, \ensuremath{m_{\rm CP}=2}, and
\ensuremath{|\mathcal O|=10}.

\begin{figure*}[t]
  \centering
  \includegraphics[width=0.85\textwidth]{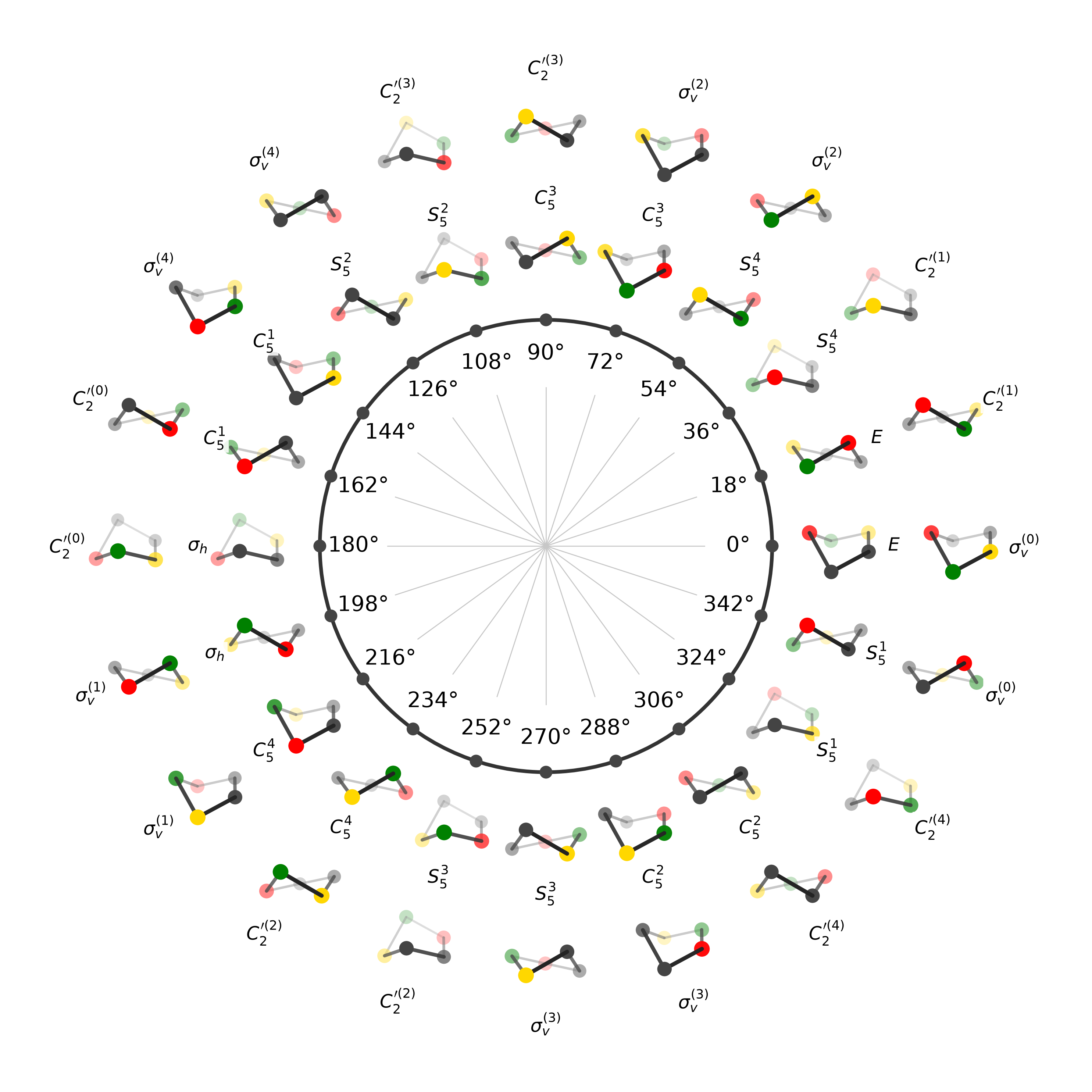}
  \caption{Envelope and twist CP phase orbits of a homogeneous
  five-membered ring at fixed puckering amplitude. Each orbit
  contains ten distinct phase points generated by twenty formal
  operations. Paired structures indicate operations producing
  the same CP coordinates. Red, green, and yellow denote ring
  elevations 1, 2, and 3, respectively.}
  \label{fig:c5_phases}
\end{figure*}

Figure~\ref{fig:c5_phases} illustrates these two orbits.
The envelope images occur at \ensuremath{36^\circ} intervals
starting from \ensuremath{0^\circ}, and the twist images at
the same intervals starting from \ensuremath{18^\circ}.
Thus, the two patterns have equal CP multiplicities and orbit
sizes, but different symmetries.

\section{Symmetry-reduced search domains for CP phases}
\label{sec:domains}

The next step is to construct symmetry-reduced CP phase domains
containing at least one representative of each orbit of
symmetry-equivalent puckering patterns. For a single active phase,
the intervals derived below are fundamental domains. For multiple
active phases, we use covering domains that remove generic
symmetry redundancy but may retain equivalent points on their
boundaries. 

For \ensuremath{r\ge1}, we take \ensuremath{m_1}
as the leading mode. The leading-phase interval is determined from the symmetry
images of that isolated mode, allowing the remaining phases
and, when present, the special-coordinate sign to change
under the same operation.

The secondary phases can then be restricted sequentially.
At each step, only symmetry operations that leave all
previously restricted phases unchanged are used to reduce
the next phase. Their intervals therefore depend on the
ordered partial sets of active modes and cannot be chosen
independently. When the special coordinate is active,
its sign must also be included in the reduction. In general, imposing
a positive sign from the outset generally requires a
different phase domain.

\subsection{Homogeneous rings}
\label{app:stepsize}

\subsubsection{Reduced interval with one active mode}

We first consider a single active phase-bearing mode,
with all other amplitudes, including the special coordinate
when present, set to zero. The symmetry reduction therefore
depends only on the images of the phase
\ensuremath{\phi_{m_1}}. 

Cyclic rotations and global inversion generate phase
equivalences that hold throughout the phase interval,
rather than only at particular geometries. Their associated
redundancy factors are denoted by \ensuremath{f_{\rm R}^{(1)}} and the global inversion,
\ensuremath{f_{\rm I}^{(1)}}, respectively. The superscript refers here to the first
and only active mode. Reflection equivalences are considered
separately below.  For a homogeneous ring, the rotational redundancy factor is
\begin{equation}
f_{\rm R}^{(1)}=\gcd(N,m_1).
\end{equation}
This factor gives the number of cyclic rotations that reproduce the
original phase, including the identity. Since there are
\ensuremath{N} possible cyclic rotations and each distinct phase is
generated \ensuremath{f_{\rm R}^{(1)}} times, the rotational phase
spacing satisfies
\begin{equation}
N\Delta_{\rm R}=2\pi f_{\rm R}^{(1)}
\end{equation}
and therefore
\begin{equation}
\Delta_{\rm R}
=
\frac{2\pi f_{\rm R}^{(1)}}{N}.
\end{equation}

Global inversion adds a phase shift of \ensuremath{\pi}.
We set \ensuremath{f_{\rm I}^{(1)}=2} if this shift is
already generated by cyclic rotation of the isolated
leading mode, and \ensuremath{f_{\rm I}^{(1)}=1}
otherwise. Thus,
\begin{equation}\label{eq:f_I}
f_{\rm I}^{(1)}
=
\begin{cases}
2, & N/f_{\rm R}^{(1)}\ \text{even},\\
1, & N/f_{\rm R}^{(1)}\ \text{odd}.
\end{cases}
\end{equation}

Including global inversion doubles the number of available operations
from \ensuremath{N} to \ensuremath{2N} since cyclic rotations alone adds $N$ and
cyclic rotations combined with inversion adds another $N$. The combined redundancy
factor is
\begin{equation}\label{eq:f_RI_1}
f_{\rm RI}^{(1)}= f_{\rm R}^{(1)}f_{\rm I}^{(1)}= \gcd(2m_1,N)
\end{equation}
The phase spacing after including rotations and global
inversion satisfies
\begin{equation}
2N\Delta_{\rm RI}
=
2\pi f_{\rm R}^{(1)}f_{\rm I}^{(1)}
\end{equation}
and therefore
\begin{equation}\label{eq:RI_period}
\Delta_{\rm RI}
=
\frac{\pi f_{\rm RI}^{(1)}}{N}.
\end{equation}

Reflection equivalences fold this phase sector in half.
The resulting interval for the leading phase is given by
\begin{equation}\label{eq:funddom_general}
\mathcal F_{m_1}
=
\left[
0,
\frac{\pi f_{\rm RI}^{(1)}}{2N}
\right].
\end{equation}
A useful example is cyclopentane. Here \ensuremath{N=5} and the
puckering is described by the single mode \ensuremath{m=2}. Since
\ensuremath{N} is odd, the \ensuremath{\pi} shift generated by the
improper block cannot coincide with a rotational image. Thus Type-I does
not hold and \ensuremath{f_{\rm I}=1}. Moreover,
\ensuremath{f_{\rm R}=1}. The combined proper and improper
blocks therefore give an effective period of \ensuremath{36^\circ}.
This is not yet the fundamental domain, because reflection symmetry
folds this sector in half:
\begin{equation}
\mathcal{F}_2
=
\left[
0,
\frac{\pi}{10}
\right]
=
[0^\circ,18^\circ].
\end{equation}
The two envelope forms at \ensuremath{0^\circ} and \ensuremath{36^\circ}
are related by the symmetry of the homogeneous ring, while the twist
form at \ensuremath{18^\circ} lies at the midpoint of the sector.
Therefore the interval \ensuremath{[0^\circ,18^\circ]} already contains
one representative of the single-mode symmetry-inequivalent
conformations of cyclopentane.

As a second example, consider cyclohexane with active mode
\ensuremath{m=2}. In this case \ensuremath{f_{\rm R}=2} and
\ensuremath{f_{\rm I}=1}, giving
\begin{equation}
\mathcal{F}_2
=
\left[
0,
\frac{\pi}{6}
\right]
=
[0^\circ,30^\circ].
\end{equation}

The corresponding one-mode intervals for rings with
\ensuremath{N=5}--\ensuremath{20} are listed in
Table~\ref{tab:funddom}. They are also useful as the base for the step size in multi-mode search
as described in Section~\ref{sec:search}

\begin{table}[ht]
\centering
\caption{Fundamental phase intervals for homogeneous rings with one
active CP phase, reported in degrees. Only modes with an associated
independent phase variable are included. Modes with identical reduction
factors and identical fundamental intervals are grouped in the same row.
The intervals correspond to the fully reduced single-phase domains,
including the final folding by reflection symmetry.}
\label{tab:funddom}
\begingroup
\renewcommand{\arraystretch}{1.18}
\begin{tabular}{ccccc|ccccc}
\hline
$N$ & $m$ & $f_{\rm R}$ & $f_{\rm I}$ & $\mathcal{F}_m$
& $N$ & $m$ & $f_{\rm R}$ & $f_{\rm I}$ & $\mathcal{F}_m$ \\
\hline
5  & any       & 1 & 1 & $[0^\circ,18^\circ]$
& 15 & $2,4,7$   & 1 & 1 & $[0^\circ,6^\circ]$ \\

6  & 2         & 2 & 1 & $[0^\circ,30^\circ]$
& 15 & $3,6$     & 3 & 1 & $[0^\circ,18^\circ]$ \\

7  & any       & 1 & 1 & $[0^\circ,12.857^\circ]$
& 15 & 5         & 5 & 1 & $[0^\circ,30^\circ]$ \\

8  & 2         & 2 & 2 & $[0^\circ,45^\circ]$
& 16 & $2,6$     & 2 & 2 & $[0^\circ,22.5^\circ]$ \\

8  & 3         & 1 & 2 & $[0^\circ,22.5^\circ]$
& 16 & $3,5,7$   & 1 & 2 & $[0^\circ,11.25^\circ]$ \\

9  & $2,4$     & 1 & 1 & $[0^\circ,10^\circ]$
& 16 & 4         & 4 & 2 & $[0^\circ,45^\circ]$ \\

9  & 3         & 3 & 1 & $[0^\circ,30^\circ]$
& 17 & any       & 1 & 1 & $[0^\circ,5.294^\circ]$ \\

10 & $2,4$     & 2 & 1 & $[0^\circ,18^\circ]$
& 18 & $2,4,8$   & 2 & 1 & $[0^\circ,10^\circ]$ \\

10 & 3         & 1 & 2 & $[0^\circ,18^\circ]$
& 18 & 3         & 3 & 2 & $[0^\circ,30^\circ]$ \\

11 & any       & 1 & 1 & $[0^\circ,8.182^\circ]$
& 18 & $5,7$     & 1 & 2 & $[0^\circ,10^\circ]$ \\

12 & 2         & 2 & 2 & $[0^\circ,30^\circ]$
& 18 & 6         & 6 & 1 & $[0^\circ,30^\circ]$ \\

12 & 3         & 3 & 2 & $[0^\circ,45^\circ]$
& 19 & any       & 1 & 1 & $[0^\circ,4.737^\circ]$ \\

12 & 4         & 4 & 1 & $[0^\circ,30^\circ]$
& 20 & $2,6$     & 2 & 2 & $[0^\circ,18^\circ]$ \\

12 & 5         & 1 & 2 & $[0^\circ,15^\circ]$
& 20 & $3,7,9$   & 1 & 2 & $[0^\circ,9^\circ]$ \\

13 & any       & 1 & 1 & $[0^\circ,6.923^\circ]$
& 20 & $4,8$     & 4 & 1 & $[0^\circ,18^\circ]$ \\

14 & $2,4,6$   & 2 & 1 & $[0^\circ,12.857^\circ]$
& 20 & 5         & 5 & 2 & $[0^\circ,45^\circ]$ \\

14 & $3,5$     & 1 & 2 & $[0^\circ,12.857^\circ]$
& \multicolumn{5}{c}{} \\

\hline
\end{tabular}
\endgroup
\end{table}

\subsubsection{Reduced interval with one active phase and the special coordinate}
\label{sec:one_phase_special}

We now consider an even-membered ring with one active
phase-bearing mode and a nonzero special coordinate,
with all remaining amplitudes set to zero.

If both special-coordinate signs are retained in the
sampling, the phase interval derived above remains
applicable, since its construction allows the sign to
change under the symmetry operations. Alternatively,
we may impose $q_{N/2}>0$
Any configuration with a negative special coordinate
can first be mapped into this sector by global inversion.
Subsequent phase reduction must then use only operations
that preserve the positive sign of the special coordinate.
Under a cyclic displacement \ensuremath{k}, this coordinate
acquires the factor \ensuremath{(-1)^k}. Rotations alone therefore
preserve its sign only for even displacements, leaving
\ensuremath{N/2} available operations. Their redundancy factor is
\begin{equation}
f_{\rm R}^{(1),+}= \gcd\left(\frac N2, m_1\right)
\end{equation}
Accordingly, the rotational phase spacing within the positive
sector satisfies
\begin{equation}
\frac N2\,\Delta_{\rm R}^{(+)}
=
2\pi f_{\rm R}^{(1),+}.
\end{equation}
Including global inversion doubles the number of sign-preserving
operations from \ensuremath{N/2} to \ensuremath{N}: even
displacements without inversion and odd displacements combined
with inversion. Defining the additional redundancy factor by
\begin{equation}
f_{\rm RI}^{(1),+}
=
f_{\rm R}^{(1),+}f_{\rm I}^{(1),+},
\end{equation}
the combined phase spacing satisfies
\begin{equation}
N\Delta_{\rm RI}^{(+)}
=
2\pi f_{\rm R}^{(1),+}f_{\rm I}^{(1),+}.
\end{equation}
Dividing by \ensuremath{N} gives
\begin{equation}
\Delta_{\rm RI}^{(+)}
=
\frac{2\pi f_{\rm RI}^{(1),+}}{N}.
\end{equation}
The numerator contains \ensuremath{2\pi}, rather than
\ensuremath{\pi}, because only \ensuremath{N} operations
preserve the special-coordinate sign, compared with
\ensuremath{2N} operations in the phase-only case.
The combined redundancy factor is
\begin{equation}\label{eq:f_RI_1plus}
f_{\rm RI}^{(1),+}= \gcd\left(N,m_1+\frac N2\right).
\end{equation}
The reference vertical reflection with axis index
zero preserves the special-coordinate sign and
folds the phase sector in half. The resulting reduced interval is
\begin{equation}
\mathcal F_{m_1}^{(+)}
=
\left[
0,
\frac{\pi}{N}f_{\rm RI}^{(1),+}
\right].
\end{equation}
This interval provides coverage with the special coordinate
restricted to positive values. It may be wider than the
phase-only interval because operations that change the
special-coordinate sign are no longer available within
the positive sector.

For cyclohexane, the ordinary mode has index two and
the special coordinate has index three. With the
special coordinate restricted to positive values,
\begin{equation}
\gcd\left(6,2+\frac62\right)=\gcd(6,5)=1,
\end{equation}
so that
\begin{equation}
\mathcal F_2^{(+)}
=
\left[0,\frac{\pi}{6}\right]
=
[0^\circ,30^\circ].
\end{equation}
Thus the phase interval remains unchanged from the
case in which the special coordinate is zero. 

For cyclooctane, consider only the ordinary coordinate
of index two and the special coordinate of index four,
with the mode of index three set to zero. Restricting
the special coordinate to positive values gives
\begin{equation}
\gcd\left(8,2+\frac82\right)=\gcd(8,6)=2,
\end{equation}
and therefore
\begin{equation}
\mathcal F_2^{(+)}
=
\left[0,\frac{\pi}{4}\right]
=
[0^\circ,45^\circ].
\end{equation}

In both examples, including the special coordinate with its sign fixed to be positive leaves the reduced phase interval unchanged, although its magnitude introduces an additional sampling coordinate. 

The methodology extends to the active-mode sets
\ensuremath{\mathcal M} and \ensuremath{\mathcal M_+}
defined in Sec.~\ref{sec:symm_search}.
The corresponding redundancy factors and phase intervals
are summarized in
Table~\ref{tbl:phase_intervals_homogeneous}.

There are additional redundancies at the boundaries, for instance, when the first phase is fixed at \ensuremath{\phi_{m_1}=0},
the residual reflection symmetry allows the second-phase interval
listed in Table~\ref{tbl:phase_intervals_homogeneous} to be halved,
with the upper endpoint of the reduced interval included.
The intervals of all subsequent phases remain unchanged.
This reduction does not modify the tabulated redundancy factors.
Additional reductions can be applied on the boundaries of the phase domain. 
However, their implementation is cumbersome and lies beyond the scope of the present discussion.

\begin{table*}[t]
\caption{Redundancy factors and reduced phase intervals for homogeneous
rings. Rows labeled by \ensuremath{r} correspond to an inactive
special mode. Rows labeled by \ensuremath{r_+} include
an active special mode, require even \ensuremath{N},
and refer to the sector \ensuremath{q_{N/2}>0}.}
\label{tbl:phase_intervals_homogeneous}
\centering

\begingroup
\renewcommand{\arraystretch}{1.8}
\setlength{\tabcolsep}{6pt}

\resizebox{\textwidth}{!}{%
\begin{tabular}{llcc}
\hline\hline
Active modes &
Redundancy factors &
First-phase interval &
Additional-phase intervals \\
\hline

\ensuremath{r=1}
&
\ensuremath{f_{\rm RI}^{(1)}=\gcd(N,2m_1)}
&
\ensuremath{
  \phi_{m_1}\in
  \left[0,\frac{\pi f_{\rm RI}^{(1)}}{2N}\right]
}
&
--- \\

\ensuremath{r_+=2}
&
\ensuremath{
  f_{\rm RI}^{(1),+}
  =\gcd\left(N,m_1+\frac N2\right)
}
&
\ensuremath{
  \phi_{m_1}\in
  \left[0,\frac{\pi f_{\rm RI}^{(1),+}}{N}\right]
}
&
--- \\
\hline

\ensuremath{r=2}
&
\ensuremath{
  f_{\rm RI}^{(2)}=\gcd(N,2m_1,m_2-m_1)
}
&
\ensuremath{
  \phi_{m_1}\in
  \left[0,\frac{\pi f_{\rm RI}^{(1)}}{2N}\right]
}
&
\ensuremath{
  \phi_{m_2}\in
  \left[0,\frac{2\pi f_{\rm RI}^{(2)}}{f_{\rm RI}^{(1)}}\right)
} \\

\ensuremath{r_+=3}
&
\ensuremath{
  f_{\rm RI}^{(2),+}
  =\gcd\left(N,m_1+\frac N2,m_2+\frac N2\right)
}
&
\ensuremath{
  \phi_{m_1}\in
  \left[0,\frac{\pi f_{\rm RI}^{(1),+}}{N}\right]
}
&
\ensuremath{
  \phi_{m_2}\in
  \left[0,\frac{2\pi f_{\rm RI}^{(2),+}}
  {f_{\rm RI}^{(1),+}}\right)
} \\
\hline

\ensuremath{r\ge1}
&
\ensuremath{
  \begin{aligned}
    &f_{\rm RI}^{(i-1)}
    =\gcd\left(N,2m_1,\{m_k-m_1\}_{k=2}^{i-1}\right)
    \\[4pt]
    &f_{\rm RI}^{(i)}
    =\gcd\left(f_{\rm RI}^{(i-1)},m_i-m_1\right)
  \end{aligned}
}
&
\ensuremath{
  \phi_{m_1}\in
  \left[0,\frac{\pi f_{\rm RI}^{(1)}}{2N}\right]
}
&
\ensuremath{
  \begin{gathered}
    \phi_{m_i}\in
    \left[0,\frac{2\pi f_{\rm RI}^{(i)}}
    {f_{\rm RI}^{(i-1)}}\right)
    \\
    i=2,\ldots,r
  \end{gathered}
} \\

\ensuremath{r_+=r+1,\quad r\ge1}
&
\ensuremath{
  \begin{aligned}
    &f_{\rm RI}^{(i-1),+}
    =\gcd\left(N,\left\{m_k+\frac N2\right\}_{k=1}^{i-1}\right)
    \\[4pt]
    &f_{\rm RI}^{(i),+}
    =\gcd\left(f_{\rm RI}^{(i-1),+},m_i+\frac N2\right)
  \end{aligned}
}
&
\ensuremath{
  \phi_{m_1}\in
  \left[0,\frac{\pi f_{\rm RI}^{(1),+}}{N}\right]
}
&
\ensuremath{
  \begin{gathered}
    \phi_{m_i}\in
    \left[0,\frac{2\pi f_{\rm RI}^{(i),+}}
    {f_{\rm RI}^{(i-1),+}}\right)
    \\
    i=2,\ldots,r
  \end{gathered}
} \\
\hline\hline
\end{tabular}%
}
\endgroup

\end{table*}

\subsection{Rings with heteroatoms}
\label{sub:intervals_hetero}

The phase-reduction argument used for homogeneous rings also applies
to rings containing heteroatoms, provided that only operations
preserving the atom-type pattern are counted. In particular,
\ensuremath{N} must be replaced by the number \ensuremath{K} of
allowed cyclic rotations, and the redundancy factors must be
evaluated within this restricted set.

If \ensuremath{K} cyclic rotations are allowed, including global
inversion gives \ensuremath{2K} rotation--inversion operations.
Of these, \ensuremath{f_{\rm RI}^{(1)}} leave the active phase
unchanged modulo \ensuremath{2\pi}. The orbit therefore contains
\ensuremath{2K/f_{\rm RI}^{(1)}} distinct phase images, and their
spacing satisfies

\begin{equation}
2K\Delta_{\rm RI}
=
2\pi f_{\rm RI}^{(1)}
\end{equation}
Hence,
\begin{equation}
\Delta_{\rm RI}
=
\frac{\pi f_{\rm RI}^{(1)}}{K}.
\end{equation}
An allowed reflection folds this interval in half. Up to an
appropriate choice of phase origin, the reduced interval is
\begin{equation}
\mathcal F_m
=
\left[
0,\frac{\pi f_{\rm RI}^{(1)}}{2K}
\right].
\end{equation}

If the special mode is also active, we first use global inversion
to choose the sector \ensuremath{q_{N/2}>0}. Since a cyclic displacement
\ensuremath{k} changes the special coordinate by the factor
\ensuremath{(-1)^k}, each allowed rotation has exactly one
sign-preserving version: without inversion for even
\ensuremath{k}, and combined with inversion for odd
\ensuremath{k}. There are therefore \ensuremath{K}, rather than
\ensuremath{2K}, available rotation--inversion operations.
Writing their redundancy factor as
\ensuremath{f_{\rm RI}^{(1),+}}, the spacing satisfies
\begin{equation}
K\Delta_{\rm RI}^{(+)}
=
2\pi f_{\rm RI}^{(1),+}.
\end{equation}
A reflection preserving the positive sector, either alone or
combined with inversion, then gives
\begin{equation}
\mathcal F_m^{(+)}
=
\left[
0,\frac{\pi f_{\rm RI}^{(1),+}}{K}
\right].
\end{equation}

\paragraph{One heteroatom or two identical non-antipodal heteroatoms.}

In these cases, no nontrivial cyclic rotation preserves the
atom-type pattern, so \ensuremath{K=1}. Rotations alone therefore
give $ f_{\rm R}^{(1)}=1$.
Without an active special mode, inversion generates the additional
phase image \ensuremath{\phi_m+\pi}. Since this image differs from
the original phase, inversion does not increase the redundancy, i.e.,
$f_{\rm I}^{(1)}=1$.
The two available operations thus generate two distinct phase
images, giving \ensuremath{\Delta_{\rm RI}=\pi}.
A reflection preserving the atom-type pattern folds this interval
in half, yielding
\begin{equation}
\mathcal F_m
=
\left[0,\frac{\pi}{2}\right].
\end{equation}
This is consistent with the residual CP symmetry of
\ensuremath{C_{2v}} type
(see Table~\ref{tbl:chartable_hetero_c2v}).

With \ensuremath{q_{N/2}>0}, inversion alone is no longer available
within the chosen sector. The only sign-preserving
rotation--inversion operation is the identity, so
$f_{\rm RI}^{(1),+}=1$, and $ \Delta_{\rm RI}^{(+)}=2\pi$.

In this case, the reduced interval becomes
\begin{equation}
\mathcal F_m^{(+)}=[0,\pi].
\end{equation}

\paragraph{Two identical antipodal heteroatoms.}

Antipodal positions are possible only for even \ensuremath{N}.
There are now \ensuremath{K=2} allowed cyclic rotations, i.e., 
the identity and the half-ring rotation, which exchanges the
identical heteroatoms.  The rotational redundancy factor is consequently
$f_{\rm R}^{(1)}=\gcd(2,m)$.

Without an active special mode, the homogeneous-ring expression
for the combined redundancy applies with two allowed rotations, and 
$f_{\rm RI}^{(1)}=\gcd(2,2m)=2$.
The four available rotation--inversion operations (in this case $K=2$) generate only two
distinct phase images:
\begin{equation}
4\Delta_{\rm RI}
=
2\pi f_{\rm RI}^{(1)}
=
4\pi.
\end{equation}
Thus \ensuremath{\Delta_{\rm RI}=\pi}, and reflection gives
\begin{equation}
\mathcal F_m
=
\left[0,\frac{\pi}{2}\right].
\end{equation}
Although the residual CP symmetry is enlarged to
\ensuremath{D_{2h}} type, its additional operations do not
provide further reduction of an isolated active phase.

With \ensuremath{q_{N/2}>0}, only two rotation--inversion
operations preserve the chosen sector, and in analogy with the case
of one active mode plus the \ensuremath{q_{N/2}} mode, the redundancy factor is therefore
\begin{equation}
f_{\rm RI}^{(1),+}
=
\gcd\left(2,m+\frac N2\right).
\end{equation}
As in the homogeneous positive-sector reduction, this factor
counts the available operations that reproduce the original
phase. Here the available set contains only two operations,
so
\begin{equation}
2\Delta_{\rm RI}^{(+)}
=
2\pi f_{\rm RI}^{(1),+}.
\end{equation}
Applying a reflection compatible with the positive sector gives
\begin{equation}
\mathcal F_m^{(+)}
=
\left[
0,\frac{\pi}{2}
\gcd\left(2,m+\frac N2\right)
\right].
\end{equation}

\section{Conformational sampling and applications}
\label{sec:search}

\subsection{Sampling strategy}\label{sec:sampling}

The symmetry-reduced covering domains defined in
Sec.~\ref{sec:domains} provide a convenient basis for a
systematic conformational search of homogeneous rings.
The practical grid includes the lower-boundary reflection
reduction described above, but retains residual symmetry
redundancy on other boundary sets.  

We sample the active phase-bearing modes in
\ensuremath{\mathcal M}. When the special mode is active,
we use \ensuremath{\mathcal M_+} and sample
\ensuremath{q_{N/2}} separately in the positive sector.
For each of these modes, we choose the phase step as
\begin{equation}
\label{eq:stepi}
\Delta\phi_{m_i}=\sup\mathcal F_{m_i}\qquad i=1,\ldots,r,
\end{equation}
Here \ensuremath{\mathcal F_{m_i}} is the isolated-mode
fundamental interval listed in Table~\ref{tab:funddom},
not the sequential interval of the multimode domain. Thus, when two modes
are active, the phase steps are $\bigl(\Delta\phi_{m_1},\Delta\phi_{m_2}\bigr)$.
An initial preconditioned search may include configurations
with one, two, or three active phase-bearing modes,
i.e., \ensuremath{\{m_1\}}, \ensuremath{\{m_1,m_2\}},
and \ensuremath{\{m_1,m_2,m_3\}}, respectively. 

To determine the number of sampled configurations,
let \ensuremath{L_i} denote the length of the sequential
interval for \ensuremath{\phi_{m_i}} in
Table~\ref{tbl:phase_intervals_homogeneous}.
When the special coordinate is active and restricted to
positive values, the corresponding lengths are denoted
by \ensuremath{L_{i,+}}.
The first phase is sampled over a closed interval, whereas the
remaining phases are sampled over half-open intervals:
\begin{equation}
	\phi_{m_1}\in[0,L_1],
	\qquad
	\phi_{m_i}\in[0,L_i),
	\qquad i=2,\ldots,r.
\end{equation}

When the first phase is zero, the residual reflection symmetry
reduces the second-phase interval to $\phi_{m_2}\in[0,L_2/2]$,
with the upper endpoint included. The intervals of subsequent
phases remain unchanged. The chosen steps divide the corresponding interval
lengths exactly, including the halved second-phase interval.

For two active modes, the number of sampled configurations is
\begin{equation}\label{eq:sm2}
S_{\mathrm{sym}}^{(2)}(m_1, m_2)
=
\frac{L_1}{\Delta\phi_{m_1}}
\frac{L_2}{\Delta\phi_{m_2}}
+
\left(
\frac{L_2}{2\Delta\phi_{m_2}}+1
\right).
\end{equation}
The first term counts configurations with a nonzero first phase,
while the second counts those on the reflection-reduced boundary.
For even $N$ and considering $q_{N/2}>0$, there are additional configurations which are given by 
\begin{equation}\label{eq:sm2+}
S_{\mathrm{sym},+}^{(2)}\left(m_1, \frac N2\right)
=
\frac{L_{1,+}}{\Delta\phi_{m_1}}+1
\end{equation}
where $L_{1,+}$ is the length of the interval for mode $m_1$ when $q_{N/2}$ is active.

For searches involving three active ordinary modes,
the second and third phases use twice their isolated-mode
phase steps only when the doubled step divides the
corresponding interval length exactly. Otherwise,
the original step is retained. We define the effective step as
\begin{equation}
	h(L,\Delta)=
	\begin{cases}
		2\Delta, & L/(2\Delta)\in\mathbb{Z}_{>0},\\
		\Delta, & \text{otherwise}.
	\end{cases}
\end{equation}

The second phase is sampled over its full interval when
the first phase is nonzero, but over an interval of half
that length when the first phase is zero. We therefore
choose its step separately in these two regions.
The doubled step may divide the full interval exactly
without dividing the halved interval exactly.
We use \ensuremath{h_2} for the full interval and
\ensuremath{h_{2,\mathrm b}} for the reflection-reduced
boundary interval, where \ensuremath{\mathrm b} denotes
the boundary. The third-phase step is denoted by
\ensuremath{h_3}. Therefore,
\begin{equation}
	\begin{aligned}
		h_2 &= h(L_2,\Delta\phi_{m_2}),\\
		h_{2,\mathrm b} &= h(L_2/2,\Delta\phi_{m_2}),\\
		h_3 &= h(L_3,\Delta\phi_{m_3}).
	\end{aligned}
\end{equation}
For example, if the full second-phase interval has a
length of sixty degrees and the original step is ten
degrees, we use a step of twenty degrees in that interval.
On the boundary, the interval has a length of thirty
degrees, so we retain the original ten-degree step.

The number of sampled configurations for three active
ordinary modes is therefore

\begin{equation}
\begin{aligned}
	S_{\mathrm{sym}}^{(3)}(m_1,m_2,m_3)
	={}&
	\Biggl[
	\left\lfloor\frac{L_1}{\Delta\phi_{m_1}}\right\rfloor
	\left\lceil\frac{L_2}{h_2}\right\rceil
	\\
	&\quad+
	\left(
	\left\lfloor\frac{L_2}{2h_{2,\mathrm{b}}}\right\rfloor+1
	\right)
	\Biggr]
	\left\lceil\frac{L_3}{h_3}\right\rceil .
\end{aligned}
\end{equation}

The first term counts points with a nonzero leading phase,
and the second counts points on its zero-phase boundary.
Floor and ceiling functions account for closed and
half-open intervals, respectively, without adding
endpoints outside the regular grid.

For even $N$ and considering $q_{N/2}>0$ the count with three modes should include
\begin{equation}\label{eq:sm3+}
S_{\mathrm{sym},+}^{(3)}\left(m_1,m_2,\frac N2\right)
=
\frac{L_{1,+}}{\Delta\phi_{m_1}}
\frac{L_{2,+}}{\Delta\phi_{m_2}}
+
\left(
\frac{L_{2,+}}{2\Delta\phi_{m_2}}+1
\right).
\end{equation}

There are $M-1$ available ordinary modes, since their
indices run from 2 to $M$. A combination of $r$ active
ordinary modes can therefore be selected in
\begin{equation}
C^{(r)}=\binom{M-1}{r}
\end{equation}
ways. For even-membered rings, the total number of combinations
with \ensuremath{r_+} active amplitudes is obtained by
adding combinations of \ensuremath{r_+} ordinary amplitudes
and combinations of \ensuremath{r=r_+-1} ordinary amplitudes
plus the special coordinate:
\begin{equation}
	C_{\mathrm{total}}^{(r_+)}
	=
	\binom{M-1}{r_+}
	+
	\binom{M-1}{r}
	=
	\binom{M}{r_+}.
\end{equation}

In addition to the phases, initial mode amplitudes must be specified for
conformational sampling. The total puckering amplitude
\ensuremath{Q\geq 0} is defined by
\begin{equation}
	Q^2=
	\begin{cases}
		\displaystyle\sum_{m=2}^{M}q_m^2, & N\ \text{odd},\\[6pt]
		\displaystyle\sum_{m=2}^{M}q_m^2+q_{N/2}^2, & N\ \text{even}.
	\end{cases}
\end{equation}
An effective strategy is to obtain \ensuremath{Q} from an optimized
reference geometry and distribute its squared magnitude equally among
the active modes. For a set \ensuremath{\mathcal M} of
\ensuremath{r=|\mathcal M|} active ordinary modes, with the special
coordinate absent or inactive, the initial mode amplitudes are
therefore
\begin{equation}
	q_m=\frac{Q}{\sqrt{r}},
	\qquad m\in\mathcal M,
\end{equation}
which preserves the total puckering amplitude.
When the special mode is also active, \ensuremath{r_+=|\mathcal M_+|=r+1}, and
the equal-amplitude initialization in the positive sector is then
\begin{equation}
	q_m=\frac{Q}{\sqrt{r_+}},
	\qquad m\in\mathcal M_+.
\end{equation}
As new conformations become available, 
the initial guess for each
subsequent structure should be constructed from the closest available
geometry, following the procedure described in Ref.~\onlinecite{AFR2026rings}.

\subsection{Local amplitude balance nomenclature}

To support conformational searches, we introduce the
semidiscrete Local Amplitude Balance (LAB) nomenclature,
which groups ring geometries according to the relative
magnitudes of consecutive puckering amplitudes.
By distinguishing absent, equal-amplitude, dominant,
and subordinate modes, this scheme provides a compact
framework for organizing sampled structures and exploring
phase variations within a given amplitude pattern.

Unlike cumulative hyperspherical descriptions, each
comparison involves only two neighboring modes, so changing
one amplitude affects at most two symbols in the classification.
For six- and seven-membered rings, there are only two puckering
amplitudes and hence a single amplitude-balance angle. In these
cases, the local and hyperspherical descriptions coincide when
the same amplitude ordering and angular convention are used.
For six-membered rings, the magnitude of the special coordinate
enters this comparison, while its sign is recorded separately.
The distinction between local and cumulative amplitude comparisons
arises only for rings with three or more puckering amplitudes.

The resulting LAB labels provide a qualitative classification
of amplitude patterns, facilitating comparisons between
conformational families while retaining the full CP coordinates
for their quantitative description.

We describe local amplitude balance through the angles
\begin{equation}
\psi_{m-2}=\operatorname{atan2}(q_{m-1},q_m),
\qquad m=3,\ldots,M.
\end{equation}
For even-membered rings, an additional angle compares the
last ordinary amplitude with the magnitude of the special
coordinate:
\begin{equation}
\psi_{M-1}
=\operatorname{atan2}(q_M,|q_{N/2}|).
\end{equation}
Thus, all local angles lie between 0 and 90 degrees.
The special coordinate is taken in absolute value for
angle calculation and symbol assignment, while its sign
is recorded separately in an external subscript.
If both amplitudes vanish, the angle is undefined and
the pair is assigned the symbol 0.

Each consecutive pair is assigned one symbol according to
Table~\ref{tab:local_nomenclature}.  For LAB assignments of optimized geometries, amplitudes
with magnitudes below \ensuremath{0.001}\,\AA\, 
are treated as zero. The phases associated with these
amplitudes are omitted from the labels.

\begin{table}[!htpb]
	\centering
	\caption{Symbols for the LAB
		nomenclature, based on the relative magnitudes of consecutive
		puckering amplitudes. For the terminal pair in even-membered
		rings, the same conditions apply to
		\ensuremath{(q_M,|q_{N/2}|)}.}
	\label{tab:local_nomenclature}
	
$$
	\begin{array}{cll}
		\hline
		\noalign{\vskip 4pt}
		\text{Symbol} & \text{Condition} & \text{Meaning}\\
		\hline
		\noalign{\vskip 4pt}
		\mathrm{Z} & q_{m-1}=0,\ q_m>0
		& \text{zenith (pole)}\\[4pt]
		\mathrm{d} & 0<q_{m-1}<q_m
		& \text{current mode dominant}\\[4pt]
		\mathrm{M} & q_{m-1}=q_m>0
		& \text{equal magnitudes}\\[4pt]
		\mathrm{s} & q_{m-1}>q_m>0
		& \text{current mode subordinate}\\[4pt]
		\mathrm{E} & q_{m-1}>0,\ q_m=0
		& \text{equator}\\[4pt]
		\mathtt{0} & q_{m-1}=q_m=0
		& \text{both modes absent}\\
		\hline
	\end{array}
$$
	
\end{table}

Z, M, and E correspond to local angles of 0, 45, and
90 degrees, respectively, and d and s denote open sectors.
Dominance (d) and subordination (s) refer only to
the current mode relative to its immediate predecessor.

The conformational label consists of a symbol string enclosed
in square brackets, followed by the ring size and the phases
of the active ordinary modes:
\begin{equation}
[\Lambda]_{(\sigma)}
(N;\phi_{m_1},\ldots,\phi_{m_r}),
\end{equation}
with
\begin{equation}
\Lambda=
\begin{cases}
	\lambda_1\cdots\lambda_{M-2},
	& N\text{ odd},\\
	\lambda_1\cdots\lambda_{M-1},
	& N\text{ even}.
\end{cases}
\end{equation}
The symbols correspond to consecutive ordinary-mode pairs,
\begin{equation}
\lambda_{m-2}\longleftrightarrow(q_{m-1},q_m),
\qquad m=3,\ldots,M,
\end{equation}
with the additional terminal symbol
\begin{equation}
\lambda_{M-1}
\longleftrightarrow(q_M,|q_{N/2}|)
\end{equation}
for even-membered rings. When the special coordinate is active,
its sign is indicated by the external subscript:
\begin{equation}
\sigma=
\begin{cases}
	+, & q_{N/2}>0,\\
	-, & q_{N/2}<0.
\end{cases}
\end{equation}

When the special coordinate vanishes, the terminal symbol
is E or 0 and the subscript is omitted.
When it is active, the external subscript indicates its sign.
The two poles correspond to opposite signs of the special
coordinate, with all ordinary amplitudes equal to zero.
No sign subscript is used for odd-membered rings.

Only phases of active ordinary modes, i.e., those with
nonzero amplitudes, are included in the label. 
The retained phases are listed in increasing
mode-index order, with their indices inferred
from the amplitude pattern. The special coordinate
has no associated phase. Empty symbol strings are omitted
together with their brackets, and empty phase lists are omitted
together with the separating semicolon.

The bracketed symbol string alone may be used as an abbreviated
label for an amplitude-pattern family, retaining the sign
subscript when the special coordinate is active.
The ring size and phase list provide the additional information
needed to distinguish its members at the LAB level.
The label describes qualitative amplitude balance but does not
encode the total puckering amplitude or replace the complete
CP coordinates. Examples are given in
Appendix~\ref{app:nomenclature_examples}.

For medium-sized and large homogeneous rings, Dale employed
a geometric nomenclature based on the arrangement of ring
segments between corners.~\cite{dale1973}
For example, the square and rectangular conformations of
cyclohexadecane are designated [4444] and [3535], respectively.
The numerical labels of Dale describe the geometric organization
of the ring, whereas LAB symbols encode local relationships
between consecutive CP amplitudes.
The two descriptions are therefore complementary rather
than interchangeable. LAB labels provide a systematic
classification of amplitude patterns, including those
of low-symmetry and mixed-mode conformations that do not
fit a simple geometric designation, and the same scheme
applies across different ring sizes.

\subsection{Application to cyclohexadecane}

The conformational search strategy outlined in
Section~\ref{sec:sampling} was applied to cyclohexadecane
(hereafter, C\ensuremath{_{16}}), a sixteen-membered cycloalkane
that provides a useful test of the present framework.
Its conformational preferences have been investigated through
strain-energy calculations,~\cite{dale1973}
molecular mechanics calculations,~\cite{allinger1980}
NMR spectroscopy,~\cite{anet1975} and vibrational
spectroscopy,~\cite{shannon1989} providing a reference for both
the lowest-energy structure and the diversity of conformations
in this flexible ring.

For the conformational search of cyclohexadecane, we used
the phase step sizes specified in Table~\ref{tab:funddom}
and calculated the numbers of sampled CP configurations
from Eqs.~(\ref{eq:sm2})--(\ref{eq:sm3+}).
Symmetry reduction, combined with selective phase-grid
coarsening for three ordinary modes, decreases these
numbers from 7696 to 297 for two active amplitudes
and from 221696 to 2023 for three active amplitudes,
including combinations involving the special coordinate.
Phase-grid coarsening provides an additional reduction
of approximately a factor of three beyond symmetry alone
in the combined sampling count.
These counts refer to sampled CP points
rather than to Cartesian starting geometries because some points
may fail to yield a valid reconstruction, whereas others
may produce several distinct geometries.
Sampling counts and total reduction factors for homogeneous
rings containing six to twenty atoms, together with a
discussion of the factors governing the savings, are
provided in Appendix~\ref{app:grid_savings}.

Cartesian geometries were generated using the ring reconstruction
algorithm developed in our previous work.~\cite{AFR2026rings}
CP coordinates specify the out-of-plane elevations but do not
uniquely determine the in-plane geometry.
Zou and co-workers~\cite{zou2020_ring_deformation_coordinates}
addressed this limitation by introducing additional coordinates
to describe in-plane deformations. Our algorithm instead
determines the projected geometry from the target elevations
and reference bond lengths by enforcing ring closure and
minimizing bond-angle distortions. The three-dimensional
structure is recovered by incorporating the target elevations,
which remain fixed along with the bond lengths during the
angular adjustment. For each sampled CP point, we explored
different arrangements of convex and concave projected bond
angles, allowing between zero and seven concave angles
(projected interior angles greater than
\ensuremath{180^\circ}). All successfully reconstructed
geometries passing the geometric checks were retained,
and redundant structures were removed before
electronic-structure optimization.

The resulting candidates were optimized at the HF/3-21G
level.~\cite{321g} Converged structures within 3.50 kcal/mol
of the lowest-energy structure found at this level were
reoptimized using the PBE0 functional~\cite{pbe0}
with the def2-TZVP basis set.~\cite{def2tzvp}
Single-point energies were subsequently calculated at the
coupled-cluster (CC) DLPNO-CCSD(T) level~\cite{dlpnoccsdt}
with the cc-pVTZ basis set.~\cite{ccpvtz} All HF and DFT calculations were performed with $Gaussian16$~\cite{g16} and all single-point CC calculations with ORCA6~\cite{ORCA6}. All this information
is available on Zenodo~\cite{rings2026}

\begin{figure}[!htpb]
	\centering
	\includegraphics[width=0.48\textwidth]{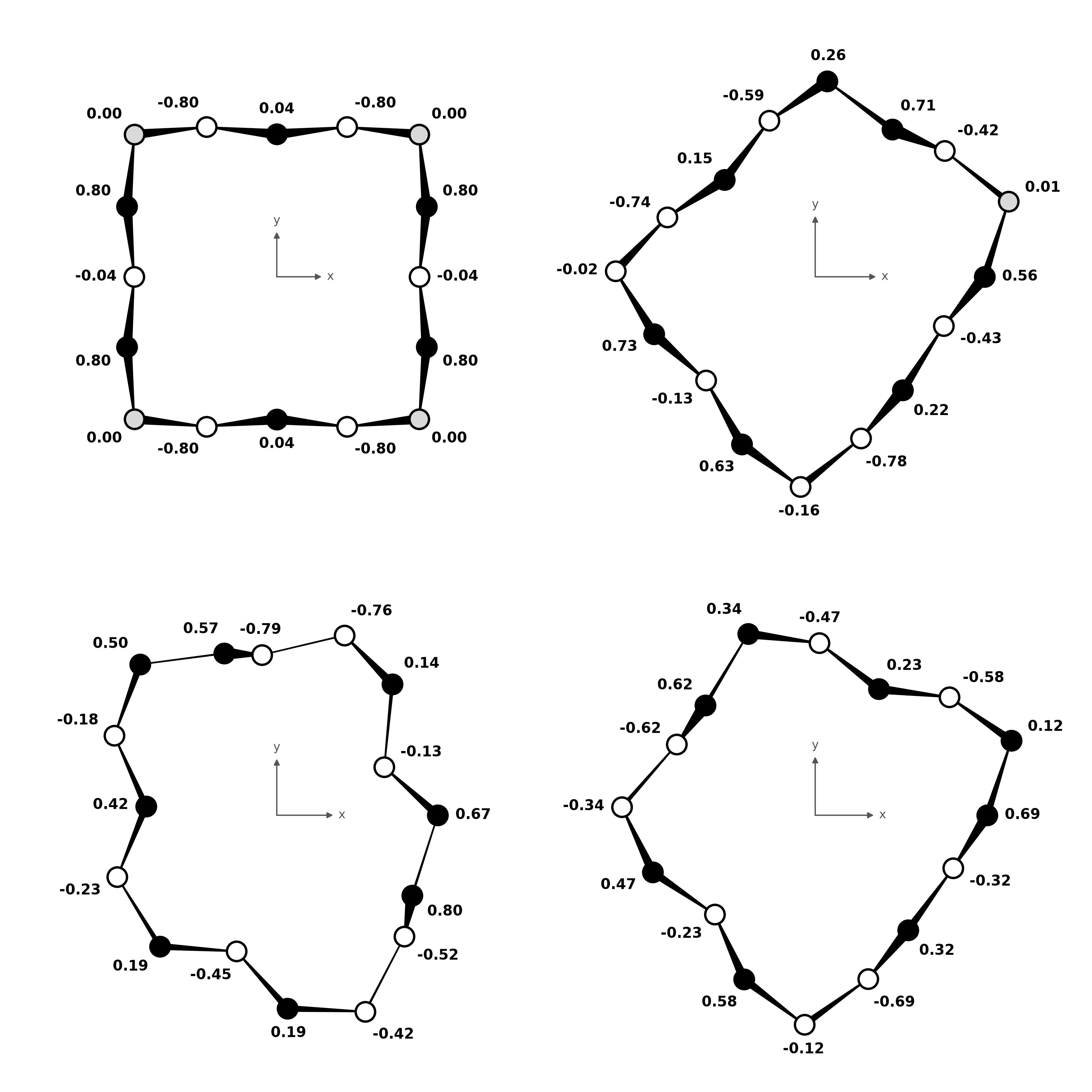}
	\caption{The four lowest-energy conformers of
		C\ensuremath{_{16}}, shown as projections onto their
		CP mean planes with elevations obtained at the DFT level.
		Top left: \ensuremath{[\mathrm{E00ZE0}]};
		top right: \ensuremath{[\mathrm{sssdss}]_{(+)}};
		bottom left: \ensuremath{[\mathrm{ssdsss}]_{(+)}};
		bottom right: \ensuremath{[\mathrm{dsddss}]_{(+)}}.}
	\label{fig:C16}
\end{figure}

Although the starting geometries were generated from CP points
with only two or three active amplitudes, the subsequent
electronic-structure optimizations were not restricted to
these modes. The search yielded 780 conformers at the
HF/3-21G level. Table~\ref{tab:c16_lowest_cc} lists
the ten lowest-energy conformers after DFT reoptimization and vibrational frequency calculations,
ranked by their DLPNO-CCSD(T) energies including zero-point
energy (ZPE) corrections at the DFT level. The four lowest-energy structures are
shown in Fig.~\ref{fig:C16} as projections onto their
CP mean planes.

\begin{table}[t]
	\centering
	\renewcommand{\arraystretch}{1.25}
	\caption{The ten lowest-energy conformers of cyclohexadecane,
		ranked by their relative CC energies including zero-point
		energy corrections. Relative DFT and CC energies, both
		including ZPE, are reported in kcal/mol with respect to
		the lowest-energy conformer at the CC+ZPE level.
		LAB labels and molecular point groups (PG)
		refer to the optimized geometries.
		Phase values in the labels are expressed in degrees,
		with the degree symbol omitted.
		Generic phases are rounded to the nearest degree, but
		symmetry-constrained values are retained at their exact
		angular values.}
	\label{tab:c16_lowest_cc}
	\begin{tabular}{lcrr}
		\hline
		LAB label & PG
		& \ensuremath{\Delta(E_{\mathrm{DFT}})}
		& \ensuremath{\Delta(E_{\mathrm{CC}})} \\
		\hline
		\noalign{\vskip 4pt}
		
		\ensuremath{[\mathrm{E00ZE0}](16;0,180)}
		& \ensuremath{D_{2d}} & 0.00 & 0.00 \\[4pt]
		
		\ensuremath{[\mathrm{sssdss}]_{(+)}(16;4,24,224,70,190,155)}
		& \ensuremath{C_1} & 2.05 & 2.03 \\[4pt]
		
		\ensuremath{[\mathrm{ssdsss}]_{(+)}(16;19,81,154,10,66,308)}
		& \ensuremath{C_1} & 1.75 & 2.03 \\[4pt]
		
		\ensuremath{[\mathrm{dsddss}]_{(+)}(16;22.5,79,315,191,67.5,124)}
		& \ensuremath{C_2} & 1.90 & 2.07 \\[4pt]
		
		\ensuremath{[\mathrm{dssdsE}](16;0,45,270,135,180,225)}
		& \ensuremath{C_2} & 2.24 & 2.59 \\[4pt]
		
		\ensuremath{[\mathrm{sddssd}]_{(+)}(16;20,162,318,280,68,125)}
		& \ensuremath{C_1} & 2.51 & 2.62 \\[4pt]
		
		\ensuremath{[\mathrm{EZEZEZ}]_{(+)}(16;22.5,315,67.5)}
		& \ensuremath{D_2} & 1.86 & 2.68 \\[4pt]
		
		\ensuremath{[\mathrm{ssdsss}]_{(+)}(16;14,4,211,136,294,28)}
		& \ensuremath{C_1} & 2.71 & 2.83 \\[4pt]
		
		\ensuremath{[\mathrm{ddsdsd}]_{(+)}(16;0,90,180,90,180,90)}
		& \ensuremath{C_s} & 2.84 & 3.09 \\[4pt]
		
		\ensuremath{[\mathrm{ssddss}]_{(+)}(16;0,0,180,180,180,180)}
		& \ensuremath{C_s} & 3.05 & 3.16 \\
		
		\hline
	\end{tabular}
\end{table}

The lowest-energy conformer found in our search is the square
structure, labeled
\ensuremath{[\mathrm{E00ZE0}](16;0,180)},
with molecular point-group symmetry \ensuremath{D_{2d}}.
This assignment agrees with the early strain-energy calculations
of Dale,~\cite{dale1973} which favored the square [4444]
conformation over the rectangular [3535] form.
Anet and Cheng~\cite{anet1975} subsequently supported
the predominance of [4444] at low temperature through
carbon-13 NMR spectroscopy, and the three resonances observed
in a 1:2:1 intensity ratio were consistent with its carbon
environments. Their iterative strain-energy calculations
placed [3535] 1.9 kcal/mol above [4444], whereas later
MM2 calculations by Allinger and co-workers~\cite{allinger1980}
gave a separation of approximately 3.4 kcal/mol.

In our calculations, the rectangular [3535] conformer, labeled
\ensuremath{[\mathrm{EZEZEZ}]_{(+)}(16;22.5,315,67.5)},
lies 2.68 kcal/mol above the [4444] minimum. However,
five other conformers occur between these two structures
in the CC+ZPE ranking. The two lowest of these have
\ensuremath{C_1} symmetry and both lie 2.03 kcal/mol
above the minimum at the precision reported in
Table~\ref{tab:c16_lowest_cc}. The search recovers the square structure as the lowest-energy
form and identifies several competing minima beyond the square
and rectangular conformations. This is qualitatively consistent
with the study of Shannon and co-workers,~\cite{shannon1989}
who associated the ordered low-temperature crystal with [4444]
and suggested contributions from low-symmetry conformers in the
liquid and high-temperature crystalline phases.
The change in energy ordering between the DFT and CC
results further illustrates the sensitivity of the relative
conformational energies to the computational method.

The low-symmetry minima found here show that symmetry reduction
in CP sampling does not restrict the search to symmetric
conformations. Instead, it avoids sampling equivalent descriptions
of the same ring elevation pattern, such as those obtained
by changing the starting atom or reversing the numbering
direction in a ring of equivalent atoms. This equivalence
does not imply that the corresponding molecular geometry
possesses any nontrivial point-group symmetry, and the
subsequent electronic-structure optimizations impose no
symmetry constraints.

The square conformer provides a concrete illustration.
For its reference CP point, with
\ensuremath{q_2=1.540}\,\AA\,
\ensuremath{q_6=1.641}\,\AA\,
a total puckering amplitude of \ensuremath{Q=2.250}\,\AA\, and phases
\ensuremath{(\phi_2,\phi_6)=(0,180)},
the 64 operations considered produce eight distinct
orbit points, each obtained by eight operations.
The operations yielding the same CP coordinates as
the identity are $E, C_{16}^{8}$,  $S_{16}^{4}$, $S_{16}^{12}$,
$C_2^{\prime,4}$, $C_2^{\prime,12}$, $\sigma_v^0$, $\sigma_v^8$.
These labels describe operations on the CP coordinates
and ring indexing, rather than directly naming molecular
point-group operations. For this particular geometry,
the eight operations leaving the reference CP point
unchanged were verified to correspond to the eight
operations of the independently determined molecular
point group \ensuremath{D_{2d}}. 

Representatives generating the eight distinct orbit
points are $E$, $C_{16}^{k}$ with $k=1,\ldots,7$.
They preserve the amplitudes and yield, respectively,
the phase pairs $(0,180)$, $(45,315)$, $(90,90)$, $(135,225)$, 
$(180,0)$, $(225,135)$, $(270,270)$, $(315,45)$ with phase values expressed in degrees.
Cyclic shifts separated by eight atom positions give
identical CP coordinates, while numbering reversal
and elevation-sign reversal add no further orbit points.
Thus, the 64 results comprise eight distinct CP
descriptions and 56 repetitions. All eight descriptions
belong to the same equivalence class, not to different
conformers, and only one orbit representative is needed
in the symmetric CP sampling.

Cyclohexadecane with only modes two and six active
illustrates the two-phase reduction in
Table~\ref{tbl:phase_intervals_homogeneous}.
The special coordinate is zero, so the row for two
active ordinary modes applies. Taking mode two as
the leading mode, the sequential redundancy factors are
$ f_{\rm RI}^{(1)} = \gcd(16,4)=4$, and $f_{\rm RI}^{(2)} = \gcd(16,4,6-2)=4$.
Substitution into the tabulated expressions gives
\begin{equation}
\begin{aligned}
	\phi_2 &\in
	\left[0,\frac{\pi}{8}\right]
	=[0^\circ,22.5^\circ],\\
	\phi_6 &\in
	[0,2\pi)
	=[0^\circ,360^\circ).
\end{aligned}
\end{equation}

The leading phase therefore has the same reduced interval
as the isolated mode listed in Table~\ref{tab:funddom}.
The second phase, however, must retain its full range.
On the lower boundary, where the leading phase is zero,
the residual reflection maps the second phase to its
negative, so the second interval reduces to $\phi_6\in[0^\circ,180^\circ]$
with both endpoints included.

\begin{figure}[!htpb]
	\centering
	\includegraphics[width=0.50\textwidth]{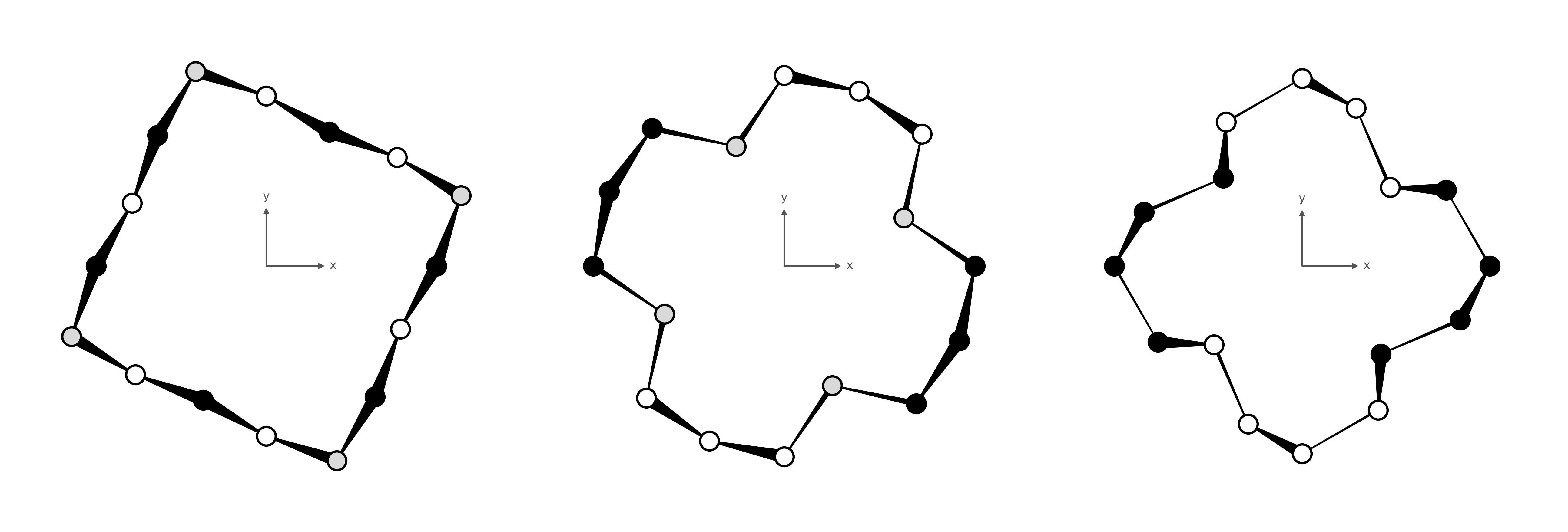}
    \caption{Three cyclohexadecane conformers sharing the LAB
	amplitude pattern \ensuremath{[\mathrm{E00ZE0}]}, shown
	as projections onto their CP mean planes.
	From left to right:
	\ensuremath{[\mathrm{E00ZE0}](16;0,180)},
	\ensuremath{[\mathrm{E00ZE0}](16;0,0)}, and
	\ensuremath{[\mathrm{E00ZE0}](16;12,61)}.
	Only modes two and six are active, but the distinct
	phase combinations produce different elevation patterns
	and projected ring shapes.}
	\label{fig:C16_d2d}
\end{figure}

The square minimum is not the only conformation with the
LAB amplitude pattern [E00ZE0].
Two additional conformers with this pattern were located
and are shown together with the minimum in
Fig.~\ref{fig:C16_d2d}.
The conformer labeled
\ensuremath{[\mathrm{E00ZE0}](16;0,0)}
lies 10.49 kcal/mol above the square minimum at the
CC+ZPE level.
The other conformer, labeled
\ensuremath{[\mathrm{E00ZE0}](16;12,61)},
has molecular \ensuremath{S_4} symmetry and lies
9.22 kcal/mol above the minimum.

These structures illustrate the conformational complexity
of large homogeneous rings. Although their puckering
involves the same two active modes, their projected ring
shapes differ substantially.
This diversity emphasizes the importance of exploring
different arrangements of convex and concave projected
bond angles during reconstruction, rather than restricting
the search to convex ring projections, where deformation coordinates 
introduced by Zou et al. can play a role.\cite{zou2020_ring_deformation_coordinates}

CP coordinates describe the out-of-plane elevations,
but do not uniquely determine the in-plane arrangement
of the atoms. Consequently, even structures with similar
CP coordinates can have markedly different Cartesian
geometries.
Symmetry makes this exploration more tractable by
restricting the phase search to a reduced domain containing
representatives of all symmetry-equivalent classes,
thereby avoiding redundant sampling without excluding
distinct CP patterns.
A comprehensive conformational search therefore combines
symmetry-reduced phase sampling with exploration of the
different projected ring geometries compatible with
each sampled CP point.

\section{Conclusions}\label{sec:conclusions}

We have derived general analytical rules that generate all
symmetry-equivalent Cremer--Pople representations of a ring puckering
pattern. These rules act directly on the phases and, for even-membered
rings, on the signed special coordinate. They allow equivalent
representations to be identified before Cartesian reconstruction
and geometry optimization.
These rules connect CP multiplicity with puckering symmetry
and provide exact single-phase intervals and reduced
multiphase search domains, with residual boundary redundancy.
The analysis also covers rings containing one heteroatom
or two identical heteroatoms.
The semidiscrete Local Amplitude Balance (LAB) nomenclature
complements this framework by organizing conformations
according to their amplitude patterns and active phases.

Symmetry makes conformational searches more efficient by
removing equivalent descriptions without excluding distinct
puckering patterns.
This reduction does not impose molecular symmetry, and the
subsequent optimizations remain unrestricted.
For cyclohexadecane, symmetry reduction alone decreases
the total number of sampled CP points by approximately
a factor of thirty. When combined with selective
phase-grid coarsening, the reduction factor approaches
one hundred.
The search recovers the square conformer as the
lowest-energy structure and identifies several competing
low-symmetry minima between the square and rectangular forms.

The conformational diversity of large rings also requires
exploration of the in-plane geometries compatible with
each CP point.
Because CP coordinates specify the elevations but do not
uniquely determine the Cartesian structure, different
arrangements of convex and concave projected bond angles
must be considered during reconstruction.
Combining this exploration with symmetry-reduced phase
sampling provides a systematic approach to ring
conformational searches while avoiding redundant
computational effort.

\section*{Autor declarations}
\subsection*{Conflict of interest}
The authors have no conflicts to disclose.

\subsection*{Author contributions}
A.F.-R. and A.L.-S. designed the method. A.L.-S. implemented the algorithm and drafted the intitial manuscript, A.F.-R. performed the electronic-structure calculations and prepared the final version of the manuscript, and both authors revised it.

\section*{Data Availability}
The data are available on Zenodo repository publicly accessible at \url{https://doi.org/10.5281/zenodo.23985089}

\appendix

\section{Classification of CP redundancies and additional examples}
\label{app:cp_redundancy}

The formal operations of a homogeneous ring form four
blocks, each containing \ensuremath{N} operations:

\begin{equation}
\begin{aligned}
	(p,s)=(0,+1)&:\quad C_N^k,\\
	(p,s)=(1,+1)&:\quad C_2^{\perp,\ell},\\
	(p,s)=(0,-1)&:\quad S_N^k,\\
	(p,s)=(1,-1)&:\quad \sigma_v^\ell.
\end{aligned}
\end{equation}

Here \ensuremath{C_2^{\perp,\ell}} denotes a twofold
rotation about an axis in the reference plane,
perpendicular to its normal.
For even-membered rings, atom and bond axes can be
distinguished by the conventional labels
\ensuremath{C_2'} and \ensuremath{C_2''}; the
corresponding mirror planes are labeled
\ensuremath{\sigma_v} and \ensuremath{\sigma_d}.

Ring-index shifts are understood modulo
\ensuremath{N}. Phase coincidences are understood
modulo \ensuremath{2\pi}, although the resulting
conditions on the original phases are conveniently
written modulo \ensuremath{\pi}.
All conditions must hold simultaneously for every
active mode.  All phases are modulo $2\pi$ and $m\in\mathcal M$. 

The symmetry interpretations below refer to the
elevation pattern on the regular reference polygon.
They need not be symmetries of a reconstructed
molecule unless its in-plane geometry is also
compatible.

\subsection{Type-R: rotational redundancies}

Type-R occurs within a single block when two cyclic shifts
generate the same CP image. A cyclic displacement $k$ transforms
the phase of each active ordinary mode according to
\begin{equation}
	\phi_m'=\phi_m+\frac{2\pi mk}{N}.
\end{equation}
Thus, two displacements $k$ and $k'$ produce a phase difference
\begin{equation}
	\Delta\phi_m=\frac{2\pi m\Delta k}{N},
	\qquad \Delta k=k-k'.
\end{equation}
The resulting CP images coincide only if this difference is
an integer number of complete turns for every active mode:
\begin{equation}
	\frac{2\pi m\Delta k}{N}=2\pi n_m,
	\qquad n_m\in\mathbb Z.
\end{equation}
Dividing both sides by $2\pi$ gives the equivalent condition

\begin{equation}
\frac{m\Delta k}{N}\in\mathbb Z.
\end{equation}

For each active ordinary mode, this requires
\ensuremath{\Delta k} to be a multiple of
\ensuremath{N/\gcd(N,m)}.
When ordinary modes are active, their simultaneous
preservation therefore gives the common
displacement step

\begin{equation}
d_{\rm R}
=
\operatorname{lcm}
\left(\frac{N}{\gcd(N,m)}\right)
=
\frac{N}{\gcd(N,\mathcal M)}.
\end{equation}
where lcm is the least common multiple. There are \ensuremath{N/d_{\rm R}} allowed
displacements modulo \ensuremath{N}, including
the identity.

If the special coordinate is active, its
transformation
\ensuremath{q'_{N/2}=(-1)^{\Delta k}q_{N/2}}
also requires an even displacement.
This is equivalent to imposing the same
divisibility condition for \ensuremath{m=N/2}.
Hence, for a nonplanar pattern,

\begin{equation}
f_{\rm R}
=
\begin{cases}
	\gcd(N,\mathcal M_+),
	& N\text{ even and }q_{N/2}\ne0,\\
	\gcd(N,\mathcal M),
	& \text{otherwise}.
\end{cases}
\end{equation}

The gcd is taken over \ensuremath{N} and all
indices in the indicated set.
If only the special coordinate is active,
this gives \ensuremath{f_{\rm R}=N/2}.
For the planar pattern, every cyclic shift
preserves the elevations and
\ensuremath{f_{\rm R}=N}.
Thus \ensuremath{f_{\rm R}} counts the cyclic
shifts preserving the complete elevation pattern,
including the identity.

\subsection{Type-C: reflected indexing with normal reversal}

Type-C connects blocks with the same
\ensuremath{s} and opposite \ensuremath{p}:

\begin{equation}
C_N^k\leftrightarrow C_2^{\perp,\ell},
\qquad
S_N^k\leftrightarrow\sigma_v^\ell.
\end{equation}

Equating the cyclic and reflected-index phases gives

\begin{equation}
\phi_m+\frac{2\pi mk}{N}
=
-\phi_m-\frac{2\pi m\ell}{N}+\pi
\end{equation}

With \ensuremath{k_{\rm C}=-(k+\ell)}, this becomes

\begin{equation}
\phi_m
=
\frac{\pi}{2}+\frac{\pi m k_{\rm C}}{N}
\pmod{\pi}.
\end{equation}

The same integer \ensuremath{k_{\rm C}} must satisfy
this condition for every active ordinary mode;
separate modewise solutions are insufficient.

Reflected indexing reverses the CP normal.
Using the reference normal, the ordered elevations
and special coordinate transform as

\begin{equation}
\begin{aligned}
	\mathcal C_\ell z_j&=-z_{N-j+\ell+2},\\
	q'_{N/2}&=(-1)^{\ell+1}q_{N/2}.
\end{aligned}
\end{equation}

If the special coordinate is active, coincidence
with the cyclic image requires
\ensuremath{(-1)^k=(-1)^{\ell+1}}.
Thus \ensuremath{k+\ell}, and therefore
\ensuremath{k_{\rm C}}, must be odd.

When the complete condition holds, an in-plane
twofold rotation preserves the elevation pattern.
Composing any one such operation with the
\ensuremath{f_{\rm R}} preserving cyclic shifts
generates all preserving Type-C operations.
Type-C therefore doubles the multiplicity supplied
by Type-R, giving \ensuremath{f_{\rm C}=2};
otherwise, \ensuremath{f_{\rm C}=1}.

\subsection{Type-I: inversion-related redundancies}

Type-I connects blocks with the same
\ensuremath{p} and opposite \ensuremath{s}:

\begin{equation}
C_N^k\leftrightarrow S_N^{k'},
\qquad
C_2^{\perp,\ell}\leftrightarrow\sigma_v^{\ell'}.
\end{equation}

Global elevation inversion adds a half-turn to
every ordinary-mode phase.
It is equivalent to a cyclic shift if a single
compensating displacement \ensuremath{k^*}
satisfies

\begin{equation}
\frac{m k^*}{N}
=
n_m+\frac12
\end{equation}
where $n_m$ is an integer number.
If the special coordinate is active, the same
condition must hold for \ensuremath{m=N/2},
or equivalently for every index in
\ensuremath{\mathcal M_+}.
For this coordinate, the condition expresses
sign compensation and requires odd
\ensuremath{k^*}.

Without an active special coordinate, a cyclic
shift with optional elevation inversion preserves
all ordinary phases when their rotational
contributions are either all integers or all
half-integers.
Choosing an active mode \ensuremath{m_1} as a
reference, the necessary and sufficient conditions
are

\begin{equation}
\frac{2m_1k}{N}\in\mathbb Z,
\qquad
\frac{(m-m_1)k}{N}\in\mathbb Z.
\end{equation}

The first condition permits either fractional
part; the difference conditions enforce the same
inversion choice for every mode.
The divisibility argument used for Type-R gives

\begin{equation}
f_{\rm RI}
=
\gcd\left(
N,2m_1,
\{m-m_1\}\right).
\end{equation}

This expression is independent of the chosen
reference mode.

If the special coordinate is active, its
preservation requires
\ensuremath{s(-1)^k=1}, hence
\ensuremath{s=(-1)^k}.
The inversion contribution is then equivalent
to \ensuremath{k/2} modulo integers, and the
ordinary-phase conditions become

\begin{equation}
\frac{(m+N/2)k}{N}\in\mathbb Z.
\end{equation}

Therefore, when ordinary modes are also active,

\begin{equation}
f_{\rm RI}
=
\gcd\left(
N,\{m+N/2\}
\right).
\end{equation}

If only the special coordinate is active,
every displacement admits exactly one preserving
inversion choice, giving
\ensuremath{f_{\rm RI}=N}.

In either nonplanar case,

\begin{equation}
f_{\rm I}=\frac{f_{\rm RI}}{f_{\rm R}}.
\end{equation}

A compensating inversion--shift operation, when
present, generates as many additional preserving
operations as there are preserving rotations.
Thus \ensuremath{f_{\rm I}=2} when Type-I holds,
and \ensuremath{f_{\rm I}=1} otherwise.
Type-I depends only on the active-mode set,
not on the phases.
For the planar pattern, elevation inversion is
trivial and \ensuremath{f_{\rm I}=2}.

\subsection{Type-P: mirror-type redundancies}

Type-P connects blocks with opposite values of
both labels:
\begin{equation}
C_N^k\leftrightarrow\sigma_v^\ell,
\qquad
C_2^{\perp,\ell}\leftrightarrow S_N^k.
\end{equation}

Elevation inversion cancels the normal reversal
associated with reflected indexing.
Equating the phases modulo $2\pi$ gives
\begin{equation}
	\phi_m+\frac{2\pi mk}{N}
	\equiv
	-\phi_m-\frac{2\pi m\ell}{N}
	\pmod{2\pi}.
\end{equation}
With \ensuremath{k_{\rm P}=-(k+\ell)}, this reduces to
\begin{equation}
	\phi_m
	\equiv
	\frac{\pi m k_{\rm P}}{N}
	\pmod{\pi}.
\end{equation}

Again, a single integer \ensuremath{k_{\rm P}}
must satisfy every active-phase condition.
The special coordinate transforms as
\ensuremath{q'_{N/2}=(-1)^\ell q_{N/2}},
so coincidence with the cyclic image requires
\ensuremath{(-1)^k=(-1)^\ell}, and hence even
\ensuremath{k_{\rm P}}, when this coordinate is active.

The complete condition identifies a mirror
operation preserving the reference elevation pattern.
Specifically, applying the inverse cyclic rotation
to the two coincident images gives a preserving mirror
with axis index
\ensuremath{a=k+\ell=-k_{\rm P}\pmod N};
this need not be the axis \ensuremath{\ell}
used in the initial comparison.
As for Type-C, one preserving operation generates
\ensuremath{f_{\rm R}} such operations by composition
with the preserving rotations.
Hence \ensuremath{f_{\rm P}=2} when Type-P is
present, and \ensuremath{f_{\rm P}=1} otherwise.

\section{Examples of the semidiscrete conformational nomenclature}
\label{app:nomenclature_examples}

Table~\ref{tab:c5_8_nomenclature} illustrates the notation for
representative conformations of six- to eight-membered rings, showing
how their local amplitude-balance angles and nonvanishing phases are encoded
in the semidiscrete labels.

For instance, the two chair conformations of cyclohexane are
naturally denoted as $[\mathrm{Z}]_{(+)}(6)$ and
$[\mathrm{Z}]_{(-)}(6)$, whereas an equatorial six-membered-ring
conformation may be labeled $[\mathrm{E}](6;\phi_2)$.
Thus, the boat conformation is $[\mathrm{E}](6;0)$ and the
skew conformation is $[\mathrm{E}](6;30)$.
In this case, the symbol E indicates that $q_2>0$ and $q_3=0$.
No sign subscript is used because the special coordinate vanishes,
and the only active phase, $\phi_2$, is retained.
The special coordinate $q_3$ has no associated phase, regardless
of whether it is active.

More generally, phases are omitted whenever their associated
ordinary-mode amplitudes vanish. For example, in a twelve-membered
ring, $\psi_1=90^\circ$ corresponds to $q_2>0$ and $q_3=0$,
so the first LAB symbol is E and $\phi_3$ is omitted from
the conformational label. The remaining symbols and phases
depend on the other active modes.

For $N=8$, if $q_2=q_3=0$ and $q_4\ne0$, the LAB symbol
string is 0Z. The first symbol is 0 because both ordinary
amplitudes vanish, whereas the terminal symbol is Z because
$q_3=0$ and $|q_4|>0$.
Both ordinary phases, $\phi_2$ and $\phi_3$, are omitted,
and the conformational label reduces to
$[\mathrm{0Z}]_{(+)}(8)$ or $[\mathrm{0Z}]_{(-)}(8)$,
depending on the sign of the surviving special coordinate $q_4$.

\begin{table}[!htbp]
	\caption{Representative conformations of six- to eight-membered rings
		expressed in terms of local amplitude-balance angles and the
		semidiscrete nomenclature. Phase values included in the labels
		are given in degrees, with the degree symbol omitted.
		A dash indicates that the corresponding local angle is undefined
		or does not exist. The subscript \ensuremath{(\pm)} denotes
		the two possible signs of the active special coordinate.}
	\label{tab:c5_8_nomenclature}
	\centering
	\renewcommand{\arraystretch}{1.25}
	\begin{tabular}{cccll}
		\hline
		\ensuremath{N} &
		\ensuremath{\psi_1} &
		\ensuremath{\psi_2} &
		Description &
		Nomenclature \\
		\hline
		6 & \ensuremath{90^\circ} & ---
		& boat (B)
		& \ensuremath{[\mathrm{E}](6;0)} \\
		6 & \ensuremath{90^\circ} & ---
		& skew (S)
		& \ensuremath{[\mathrm{E}](6;30)} \\
		6 & \ensuremath{0^\circ} & ---
		& chair (C)
		& \ensuremath{[\mathrm{Z}]_{(\pm)}(6)} \\
		\hline
		7 & \ensuremath{90^\circ} & ---
		& boat (B)
		& \ensuremath{[\mathrm{E}](7;0)} \\
		7 & \ensuremath{90^\circ} & ---
		& twist-boat (TB)
		& \ensuremath{[\mathrm{E}](7;12.857)} \\
		7 & \ensuremath{0^\circ} & ---
		& chair (C)
		& \ensuremath{[\mathrm{Z}](7;0)} \\
		7 & \ensuremath{0^\circ} & ---
		& twist-chair (TC)
		& \ensuremath{[\mathrm{Z}](7;12.857)} \\
		\hline
		8 & \ensuremath{90^\circ} & ---
		& boat-boat (BB)
		& \ensuremath{[\mathrm{E0}](8;0)} \\
		8 & \ensuremath{90^\circ} & ---
		& twist-boat (TB)
		& \ensuremath{[\mathrm{E0}](8;45)} \\
		8 & \ensuremath{0^\circ} & \ensuremath{90^\circ}
		& long-chair (L)
		& \ensuremath{[\mathrm{ZE}](8;0)} \\
		8 & \ensuremath{0^\circ} & \ensuremath{90^\circ}
		& chair (C)
		& \ensuremath{[\mathrm{ZE}](8;22.5)} \\
		8 & --- & \ensuremath{0^\circ}
		& crown (W)
		& \ensuremath{[\mathrm{0Z}]_{(\pm)}(8)} \\
		\hline
	\end{tabular}
\end{table}

\section{Reduction of phase sampling}
\label{app:grid_savings}

Table~\ref{tab:cp_grid_savings} compares the numbers of
sampled CP points in full and symmetry-reduced grids
for homogeneous rings containing six to twenty atoms.
The counts include all combinations of two or three
active amplitudes and use the original isolated-mode
phase steps in both grids.
Their difference therefore measures the reduction
due only to symmetry.
For even-membered rings, combinations involving the
special coordinate are also included.
The full grids sample both signs, whereas the
symmetry-reduced grids use only the positive sector.
Grid points follow the closed or half-open character
of each interval, without adding endpoints outside
the regular grid.

The last column of Table~\ref{tab:cp_grid_savings} lists two reduction factors,
rounded to the nearest integer and separated by a slash.
The first is obtained by dividing the combined full-grid
count for the two searches by the corresponding
symmetry-reduced count shown in the table.
The second also includes the selective increase in
phase steps described in Sec.~\ref{sec:sampling}.
For three active ordinary modes, the second and third
phase steps are doubled only when they divide the
corresponding interval lengths exactly.
This rule is applied separately to the second-phase
interval and its reflection-reduced interval on the
zero-phase boundary.
Combinations involving the special coordinate retain
the original steps.

Both factors measure reductions in sampled CP points,
not in reconstructed Cartesian geometries, optimized
conformers, or computational time.
Unlike symmetry reduction, increasing the phase steps
changes the sampling resolution and provides an
additional saving rather than removing equivalent
representations.

\begin{table}[t]
	\centering
	\caption{Full and symmetry-reduced CP grid counts for
		searches with two and three active amplitudes.
		The counts shown use the original isolated-mode phase
		steps in both grids, so their reduction is due only
		to symmetry.
		Both signs of the special coordinate are included
		in the full grids, whereas reduced grids use only
		its positive sector.
		Factor lists two rounded reduction factors for the
		combined searches, separated by a slash.
		The first includes only symmetry reduction, whereas
		the second also includes selective doubling of the
		second and third phase steps for three ordinary modes.}
	\label{tab:cp_grid_savings}
	\begingroup
	\renewcommand{\arraystretch}{1.18}
	\setlength{\tabcolsep}{5pt}
	\begin{tabular}{crrrrr}
		\hline\hline
		& \multicolumn{2}{c}{Two active amplitudes}
		& \multicolumn{2}{c}{Three active amplitudes}
		& \\
		\cline{2-3}\cline{4-5}
		N & Full & Symmetry & Full & Symmetry & Factor\\
		\hline
		6  & 24       & 2       & ---          & ---       & 12/12\\
		7  & 784      & 43      & ---          & ---       & 18/18\\
		8  & 176      & 11      & 256          & 13        & 18/18\\
		9  & 2\,160   & 93      & 15\,552      & 684       & 23/65\\
		10 & 1\,320   & 70      & 10\,400      & 408       & 25/44\\
		11 & 11\,616  & 402     & 340\,736     & 11\,792   & 29/104\\
        12 & 1\,216 & 60 & 11\,424 & 437 & 25/47\\
		13 & 27\,040  & 790     & 1\,406\,080  & 41\,080   & 34/128\\
		14 & 8\,120   & 316     & 235\,200     & 7\,157    & 33/86\\
		15 & 21\,040  & 561     & 940\,800     & 24\,392   & 39/141\\
		16 & 7\,696   & 297     & 221\,696     & 6\,591    & 33/99\\
		17 & 97\,104  & 2\,163  & 11\,005\,120 & 245\,140  & 45/173\\
		18 & 17\,832  & 560     & 838\,368     & 20\,733   & 40/117\\
		19 & 161\,728 & 3\,220  & 24\,582\,656 & 489\,440  & 50/195\\
		20 & 18\,816  & 674     & 939\,200     & 21\,767   & 43/111\\
		\hline\hline
	\end{tabular}
	\endgroup
\end{table}

The symmetry reduction depends on ring size,
the active-mode combinations, and the treatment
of domain boundaries.
Common divisors of the ring size and mode-index
combinations determine the redundancy factors
and sequential phase intervals.
Some active-mode subsets admit nontrivial operations
that leave every pattern in the subset unchanged.
These operations reduce the number of distinct
CP images per pattern and therefore the reduction
available within that subset.
For prime ring sizes, such phase-independent
redundancies are absent for subsets of ordinary modes,
giving a more uniform reduction across combinations.
For even-membered rings, the special coordinate
introduces an additional dependence on the operations
that preserve its sign.
The finite-grid reduction also reflects the relative
contribution of boundary points.

A larger reduction factor does not necessarily imply
a smaller search.
For example, symmetry alone gives rounded combined
factors of 50 and 33 for the nineteen- and
sixteen-membered rings, respectively.
Nevertheless, their symmetry-reduced grids contain
492\,660 and 6\,888 points.
The larger ring has more available mode combinations
and finer isolated-mode phase steps, resulting in
a substantially larger grid.

Selective phase-grid coarsening increases these factors
to approximately 195 and 99 and reduces the corresponding
counts to 126\,644 and 2\,320 points.
Thus, the two factors distinguish the saving obtained
by removing symmetry-equivalent descriptions from the
additional saving obtained by sampling less finely.
The reduced point count indicates the size of the
remaining sampling task.

\begin{acknowledgments}
This work has received financial support from MICIU/AEI/PID2025-173245NB-I00, Xunta de Galicia (Centro singular de Investigaci\'{o}n de Galicia, accreditation 2023-2027, ED431G 2023/03 and the European Regional Development Fund (ERDF), and Grupo de Referencia Competitiva, ED431C 2025/06) and María de Maeztu Excellence Accreditation CEX2025-001612-M funded by MICIU/AEI /10.13039/501100011033. The authors thank Galician Supercomputing Center (CESGA) for providing access to their computing facilities.
\end{acknowledgments}


%

\end{document}